\documentclass[aps,superscriptaddress,twocolumn,amsmath,amssymb]{revtex4-2}
\usepackage{graphicx}
\usepackage{epstopdf}
\usepackage{bm}% bold math
\usepackage{subfigure}
\usepackage{sidecap}

\usepackage{graphicx}
\usepackage{color}
\usepackage{epstopdf}
\usepackage{amssymb}
\usepackage{amsmath}

\usepackage{bm}% bold math
\graphicspath{{Figures/}}
\usepackage{subfigure}
\usepackage{sidecap}

\newcommand{\si}{\sigma}

\newcommand{\ga}{\gamma}
\newcommand{\eps}{\epsilon}

\newcommand{\de}{\delta}
\newcommand{\De}{\Delta}

\newcommand{\be}{\begin{equation}}
\newcommand{\ee}{\end{equation}}

\newcommand{\bea}{\begin{eqnarray}}
\newcommand{\eea}{\end{eqnarray}}
\newcommand{\bd}{\begin{displaymath}}
\newcommand{\ed}{\end{displaymath}}
\newcommand{\ba}{\begin{array}}
\newcommand{\ea}{\end{array}}
\newcommand{\bi}{\begin{itemize}}
\newcommand{\ei}{\end{itemize}}
\newcommand{\bc}{\begin{center}}
\newcommand{\ec}{\end{center}}
\newcommand{\bfl}{\begin{flushleft}}
\newcommand{\efl}{\end{flushleft}}
\newcommand{\bfr}{\begin{flushright}}
\newcommand{\efr}{\end{flushright}}
\newcommand{\non}{\nonumber}

\newcommand{\bl}{\begin{aligned}}
\newcommand{\el}{\end{aligned}}

\newcommand{\hh}{\hat{h}}

\newcommand{\hV}{\hat{V}}

\newcommand{\hde}{\hat{\delta}}

\newcommand{\fs}{\frac{1}{2}}

\newcommand{\om}{i\omega_n}

\newcommand{\ra}{\rangle}
\newcommand{\la}{\langle}

\newcommand{\GR}{GdRu$_2$Si$_2$}

\newcommand{\BLU}{\color{black}}

\def\br{{\bf r}}\def\bR{{\bf R}} 
\def\bk{{\bf k}} \def\bK{{\bf K}} \def\bq{{\bf q}}

\def\bQ{{\bf Q}}   
  
 \def\bd{{\bf d}} \def\bS{{\bf S}} 
 \def\bS{{\bf S}}  \def\bs{{\bf s}}

\def\bs{{\bf s}}

\def\da{\downarrow} \def\ua{\uparrow}

\def\={\!\!\!&=&\!\!\!}
\def\+{\!\!\!&&\!\!\!+~}
\def\-{\!\!\!&&\!\!\!-~}

\usepackage{color}
\usepackage[dvipsnames]{xcolor}
\usepackage{xcolor}
\usepackage[colorlinks=true,citecolor=blue]{hyperref}

\usepackage{hyperref,cleveref}
\begin{document}

\title{Image of helical local moment magnetic order in the STM spectrum}
%of \GR}

%
\author{Alireza Akbari}
\affiliation{Beijing Institute of Mathematical Sciences and Applications (BIMSA), Huairou District, Beijing, 101408, China}
%\email{alireza@bimsa.cn}

%
\author{Peter Thalmeier}\email[Corresponding author: \vspace{-3pt}]{peter.thalmeier@cpfs.mpg.de}
\affiliation{Max Planck Institute for the  Chemical Physics of Solids, D-01187 Dresden, Germany}
%\email{thalm@cpfs.mpg.de.cn}

\date{\today}

\begin{abstract}
The surface tunneling microscope (STM) method probes the conduction electron spectrum
which is influenced by the presence of collective order parameters. It
may in fact be used as a tool to obtain important information about their microscopic nature, for example the gap symmetry
in unconventional superconductors. Surprisingly it has been found that the STM spectrum can also 
identify  magnetic order of completely localised electrons e.g. incommensurate helical structure of 4f electron moments,
as observed in the compound \GR.
This is due to the fact that the exchange coupling of conduction states to the  localised subsystem reconstructs
the conduction bands which then leaves an imprint on the STM spectrum. We develop a theory based on this idea
that shows firstly the appearance of STM satellite peaks at the wave vector of localised moment helical order for the
pure surface. Secondly we derive the quasiparticle interference spectrum in Born approximation due to the presence of surface impurities
which contains information on the reconstruction process of itinerant states caused by the localised helical order.
Furthermore we show that within full T matrix approach  impurity bound states are also influenced by the exchange coupling to helical magnetic order.
\end{abstract}
%\pacs{ }
\maketitle

\section{Introduction}
\label{sec:intro}

The scanning surface tunneling microscopy (STM) has been successfully used to unravel the
itinerant electron characteristics of correlated metals, in particular for quasi-2D materials.
The single-site scattering from random impurities and defects on the surface leads to oscillatory
electronic density modulations due to quasiparticle interference (QPI) which are superposed on 
the background density of the periodic lattice. The QPI Fourier transform gives direct information
on the momentum structure of the conduction bands at constant energy (constant bias voltage), in particular on the Fermi surface (at zero bias voltage). This method is complementary
to ARPES experiments and has much higher energy resolution. The STM-QPI method enables the  investigation of
ordered states of materials, such as superconductors and itinerant charge (CDW) and spin (SDW) density wave states of  conduction
electrons.  Their appearance  changes the observed spectral functions by reconstruction of quasiparticle states
due to the presence of gap functions associated with the order parameters. In fact it is possible to determine the symmetry of unconventional
superconducting gap functions for small $T_c$ materials not accessible to ARPES, such as has been demonstrated in the d-wave case  of heavy fermion compound CeCoIn$_5$ \cite{akbari:11,allan:13,zhou:13} and Fe-pnictide superconductors \cite{hoffman:11,allan:12,kreisel:16,hanaguri:18} . Furthermore the QPI method can also be used to investigate superconducting (SC) finite-momentum pairing like the pair density wave state \cite{gu:23} of unconventional superconductors or the Fulde-Ferrell-Larkin-Ovchinnikov (FFLO) state in external field \cite{akbari:16}. It is also useful to study non-SC pair states of conduction electrons  as in itinerant CDW materials\cite{arguello:15} and SDW compounds \cite{kamble:16,goyal:25, Akbari:10a} and furthermore application  to altermagnets \cite{hu:24} has been proposed.

Surprisingly it has recently been found that the STM-QPI method is even applicable to investigate 
helical magnetic structures of completely localised moment origin due to strongly bound 4f- electrons
that cannot directly contribute to the tunneling current. This was observed in the tetragonal (D$_{4h}$) helical magnet
\GR~with Gd-terminated surfaces \cite{spethmann:24,yasui:20}. The localised $S=7/2$ spins of the S-state ion Gd$^{3+}$ in this compound are known to order in a bulk multi-Q structure \cite{wood:23} that contains a helix with incommensurate (IC) propagation vector $\bQ$ oriented along the tetragonal axis with $Q\simeq 0.4(\pi/a)$. Unlike for itinerant SDW compounds the strongly bound 4f electrons do not contribute to the tunneling current, nevertheless the image of this helix structure was found in the STM spectrum.
This finding opens the way to investigate peculiar surface magnetic structures that are not accessible to neutron diffraction.

In this work we take these experiments as a motivation to consider the general theoretical problem how a localised background helical magnetic order influences the STM spectra of itinerant electrons. We consider a tight binding conduction band model that contains the typical
nesting possibilities with incommensurate as well as commensurate nesting vectors leading to helical magnetic order via the associated effective intersite exchange interactions. We employ a simple magnetic structure model consisting of a single-\bQ~transverse helix for the local moments.
Thereby the STM image  of the local moment helix is generated in two steps:\\
First, the contact exchange interaction of ordered Gd $S=7/2$ spins on the lattice with conduction electron spins leads to reconstructed conduction electron bands. Already for the pure surface this causes supplementary modulations of the spectral conduction electron density with the wavelength corresponding to the localised helix moments. In the Fourier transform of  the real-space density modulations this leads to the appearance of additional satellite positions at the propagation vector of the helix which can be determined in this way, akin to the procedure in diffraction experiments.\\
 Secondly these reconstructed quasiparticles scatter from the surface impurities and defects. The Fourier transform of the nonperiodic QPI oscillations contains important information how the quasiparticle bands are reconstructed by the underlying local moment helix.
And finally we will also consider the effect of helical order on possible bound or resonance states at the impurity 
sites in full T matrix approach to the impurity scattering.
 
The theoretical foundations and analysis of these topics will be presented in detail in the following sections based on the simple
tight binding and helix model for itinerant and localised subsystems. To stay consistent the helix propagation vector used in the QPI theory is determined 
from the nesting properties of the tight binding (TB) Fermi surfaces by identifying the positions of maxima in the static Lindhard function and hence in the effective RKKY interaction. In fact this mechanism was proposed \cite{bouaziz:22,eremeev:23} for \GR~to be important although an alternative scenario based on frustration effects \cite{nomoto:20} has also been considered.

For calculation of the pure crystal topographic spectrum and the impurity induced STM-QPI images we use established local Green's function approach including the scalar and exchange impurity scattering within Born approximation. Predictions for the satellite peaks of the pure system and spectral functions of QPI images resulting from surface impurity scattering are made. They are discussed for typical Fermi surface models with different nesting properties and as function of the coupling strength between itinerant and localised components. Likewise we will investigate the influence of helical background order on bound state formation possible for strong scattering at impurity sites.

\section{The 2D electronic tight binding model 
%for G\lowercase{d}R\lowercase{u}$_2$S\lowercase{i}$_2$
and its nesting properties}
\label{sec:model}

As a generic basis for the helix state QPI calculations we employ a model composed of simple  tight binding (TB) on a {\BLU 2D square lattice with lattice constant $a \equiv 1$} and localised 4f spins $S$ on the lattice sites. {\BLU We include hopping elements up to third nearest neighbors which allows to model Fermi surfaces (FS) with various nesting properties and hence local moment ordering vectors. We want to be able to study three such FS models with strongly different nesting properties and magnetic structures  and compare their respective influence on the scanning STM spectrum.} Although this TB model is standard we briefly define it here for easy reference. The conduction bands are then given by the TB expression  
\bea
\bl
\eps_\bk= &-2t(\cos k_x+\cos k_y) -4t'\cos k_x\cos k_y
\\
& -2t'' (\cos2k_x+\cos2k_y),
\label{eq:TBband}
\el
\eea
where $t,t',t''$  are the first, second and third nearest neighbor (NN) hopping energies, respectively.
We assume a dominating $t>0$ first neighbor hopping. This model has the following energies at the symmetry 
points of the {\BLU 2D square lattice Brillouin zone (BZ) defined by  $-\pi\leq k_x,k_y\leq\pi$} : $\Gamma(0,0)$  $\eps_\Gamma=-4(t+t'+t'')$, $X(\pi,0);(0,\pi)$: $\eps_X=4(t'-t'')$ and $M(\pi,\pi): \eps_M=4(t-t'-t'')$. For $t>0; t'>-2t$ (band bottom at $\Gamma$) the bandwidth is $2D=\eps_X-\eps_\Gamma=8t+4t'$ $(t'>0)$ or $2D=\eps_M-\eps_\Gamma=8t$ $(t'<0)$. The chemical potential varies between band bottom $\eps_\Gamma$ and top of the band at $\eps_X$or $\eps_M$. 

\begin{figure}
%\begin{figure*}
%\includegraphics[width=0.25\textwidth]{FSpara}\hspace{0.7cm}
%\includegraphics[width=0.25\textwidth]{FSint}\hspace{0.7cm}
%\includegraphics[width=0.25\textwidth]{FSnest}\\
%\hspace{0.5cm}
%\includegraphics[width=0.30\textwidth]{chipara-q}
%\includegraphics[width=0.30\textwidth]{chiint-q}
%\includegraphics[width=0.30\textwidth]{chinest-q}
\includegraphics[width=\columnwidth]{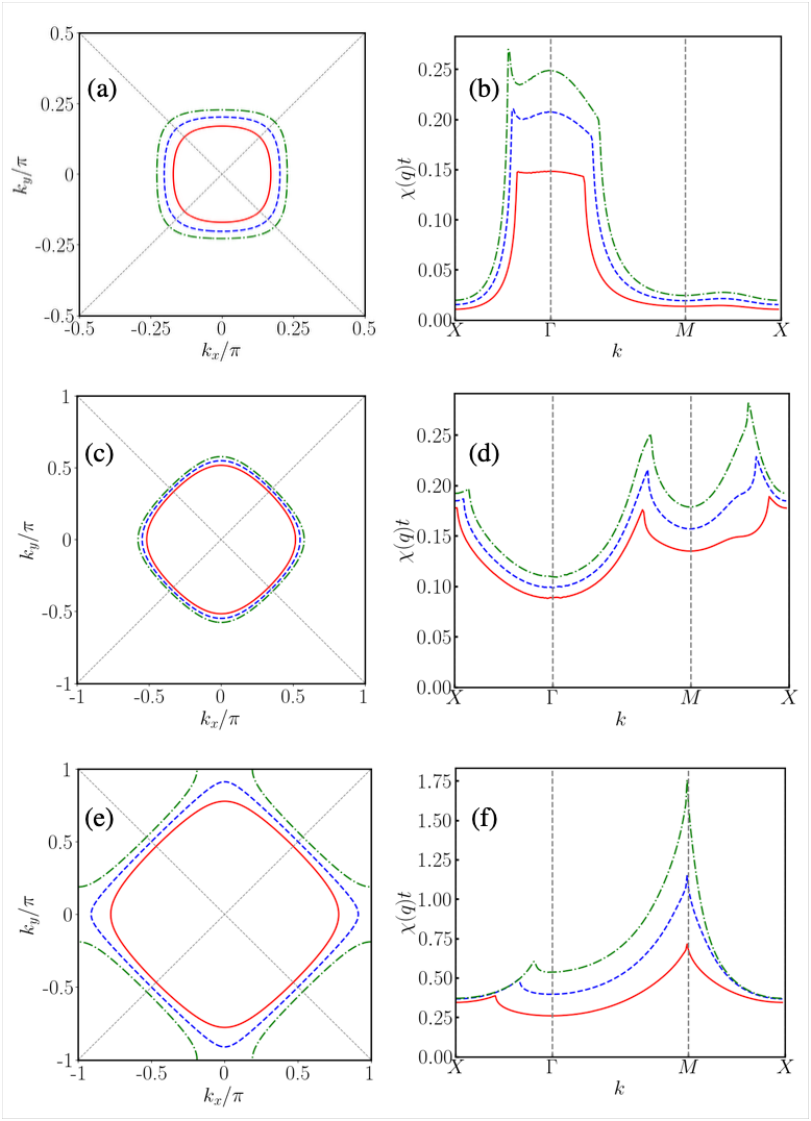}
\caption{{\it Left column}: FS sheets corresponding to TB expressions in Eq.~(\ref{eq:TBband}), using parameters {\bf (a)}  $t=1.,t'=0.8,t''=-0.55$ and Fermi energies $\eps_F/\eps_\Gamma=0.96,0.94,0.92$ consecutively for black, red, blue sheets. For this case (note the half BZ scale) the nesting vector with $\bQ\simeq (0.4\pi,0)$ corresponds approximately to the case in Ref.~\cite{bouaziz:22}. {\bf (c)}: $t=1.,t'=0.7,t''=0.5$ and Fermi energies $\eps_F/\eps_\Gamma=0.20,0.15,0.11$ with dominant nesting vector $\bQ\simeq(\pi,0.3\pi)$ {\bf (e)} $t=1.,t'=0.02,t''=0$ and Fermi energies $\eps_F/\eps_\Gamma=0.1,0.,-0.1$ with nesting vector $\bQ\simeq(\pi,\pi)$ ; both in same color order as (a). {\it Right column}: Related wave vector-dependent susceptibilities $\chi({\bq})$ for {\bf (b)} $\eps_F/\eps_\Gamma=0.94$ and $t''=-0.4, -0.5, -0.55$ in increasing sequence {\bf (d)} parameters like (c). {\bf (f)} close to perfect nesting $\epsilon_F=0$ and $t'=0.32,0.08,0.02$, both in same color order as (b).}
\label{fig:FSpara}
%\end{figure*}
\end{figure}

{\BLU For the small FS model of Fig.~\ref{fig:FSpara}(a) it is useful to consider a small momentum expansion
because it illuminates the constraint on hopping parameters to achieve axis-parallel nesting vector \bQ~such as appropriate
 for \GR.
In this  case the chemical potential is close to the bottom of the band and then the dispersion around the $\Gamma$ point} may be approximated, up to fourth order by
\be
\eps_\bk=\eps_\Gamma+\eps_2(k_x^2+k_y^2)-\eps_4(k_x^4+k_y^4)-\eps'_4k_x^2k_y^2
,
\label{eq:Gband}
\ee
where $\eps_2=t+2t'+t''$, $\eps_4=\frac{1}{3}(\frac{1}{4}t+\fs t'+4t'')$ and $\eps'_4=t'$. 
Then  $\eps_\bk =\eps_F$ determines the Fermi surface where $\eps_F$ is the Fermi energy. Due to the fourth order term the 
length of the Fermi wave vector $k_F$ depends on the direction and that determines the nesting properties discussed below. Starting from the spherical case (only second order term) with $k_F^0=(\hat{\eps}_F/\eps_2)^\fs$ $(\hat{\eps}_F=\epsilon_F-\epsilon_\Gamma$ is counted from the band bottom) we have for the two symmetry directions 
\bea
\bl
k_F^{(10)}
&\simeq
 k_F^0\bigl[1+\fs(\frac{\eps_4}{\eps_2})k_F^{02}\bigr],
\\
 k_F^{(11)}
 &\simeq
  k_F^0\bigl[1+\frac{1}{4}(\frac{\eps_4}{\eps_2}+\fs\frac{\eps'_4}{\eps_2})k_F^{02}\bigr].
 \label{eq:Fermi}
 \el
\eea
If there should be a nesting e.g. in the $(10)$ direction relatively flat FS sheets perpendicular to $(10)$ and a wave vector $Q\simeq 2k_F^0$ apart must be present. This happens when $k_F^{(10)}<k_F^{(11)}$ which requires the condition $t'>\frac{1}{4}(t+16t'')$ according to Eq.(\ref{eq:Fermi}). \\

A more quantitative and precise way to look at the nesting properties is obtained from numerical calculation of the Lindhard
function, i.e. static susceptibility for the TB bands as given by
\be
\bl
\chi(\bq)=2\sum_\bk
\frac{f_{\bk+\bq} -f_\bk}{\eps_\bk-\eps_{\bk+\bq}}.
\label{eq:sus}
\el
\ee
The nesting vector \bQ~ of the Fermi surface appears as a peak of $\chi(\bq)$
in the Brillouin zone (BZ). This function also determines the effective RKKY interaction
between the localised moments. The latter results from the coupling $I_{ex}$
of the localised  (e.g. Gd)  spins $\bS(\bR_i)$ with conduction electron
 spins $\bs(\br)$ defined below. In terms of Fourier components the RKKY interaction is given by
 % which is assumed as an isotropic
%contact exchange interaction $h_{ex}(\bR_i)=I_{ex}\bs(\br)\cdot\bS(\bR_i)\delta(\br-\bR_i)$. 
%
\bea
H_{ex}=-\fs\sum_\bq J(\bq)\bS_\bq\bS_{-\bq}
\label{eq:RKKY}
\eea
with $J(\bq)=I_{ex}^2\chi(\bq)$. The (positive) maximum of $J(\bq)$ at $\bq=\bQ$ due to nesting property of the FS then defines the magnetic ordering vector.\\
% at the N\'eel temperature $T_N=J(\bQ)S(S+1)/3k_B$.\\ 

Now we discuss briefly to which extent this simple TB model and nesting mechanism can be applied 
to the conduction bands and magnetic order in \GR. The density functional band structure calculations yield a complicated
multi-sheet structure  \cite{bouaziz:22,eremeev:23} of the FS  and this is indeed found in dHvA experiments \cite{matsuyama:23}. It was, however, proposed that the ordering vector $\bQ\simeq (0.4\pi,0)$ of the underlying helical structure of Gd spins  corresponds well to the peak in the static susceptibility $\chi(\bq)$ at the same wave vector. The latter appears
naturally as a nesting vector of a barrel-shaped quasi-2D electron sheet around the $\Gamma$ -point  \cite{bouaziz:22} . The simple model discussed above is adequate to describe this situation. 
Since the spiral ordering wave vector of Gd spins is oriented along the tetragonal axes the nesting vector should
also point in this direction which is the reason for using fourth order terms to flatten the model FS appropriately
perpendicular to \bQ. This leads indeed to the calculated (Eq.(\ref{eq:sus})) susceptibility exhibiting the maximum at the nesting wave vector (right column of Fig.~\ref{fig:FSpara}). It should be noted that for the spherical 2D FS (neglecting fourth order terms in the approximate dispersion of Eq.~(\ref{eq:Gband}) there is no peak but $\chi(\bq)$ is flat (constant) for all $|\bq|<2k^0_F$ (see Fig.~\ref{fig:FSpara}(b) at smallest Fermi energy). {\BLU We mention that a more advanced tight binding approach to model  the electronic structure for detailed correspondence to real compounds like \GR~requires the mapping of density functional band structure to a suitable basis of Wannier orbitals as has been, e.g., carried out for Fe-pnictide compounds \cite{kreisel:16}.}
%
% %%%%%%%%%%%%%%%%%%%%% figure %%%%%%%%%%%%%%%%%%%%%%%%%%%%
%%%%%%%%%%%%%%%%%%%%%%fig%%%%%%%%%%%%%%%%%%%%%%%%%%%%%%%
%

\section{Reconstructed bands and Green's functions in the spiral phase}

The helical order parameter of localised S-state ion Gd$^{3+}$
with a large quasi-classical spin $S=7/2$ may be described by a modulated
spin expectation value at site $\bR_i$ given as
\bea
\la \bS_i\ra=S({\bf a}\cos\bQ\cdot\bR_i+{\bf b}\sin\bQ\cdot\bR_i),
\label{eq:SDW}
\eea
where ${\bf a}$, ${\bf b}$ are the cartesian unit vectors in the plane and $\bQ$
is the generally incommensurate wave vector of the helix. {\BLU In the FS models
of Fig.~\ref{fig:FSpara} there are two symmetry equivalent nesting vectors 
$\bQ=(0.4\pi,0)$ or $(0,0.4\pi)$, rotated
by $\pi/2$. In Eq.~(\ref{eq:SDW}) we use for simplicity  a single-\bQ~structure corresponding 
to two possible domains with ordering vectors parallel to one of the momentum space axes,
corresponding to  Fig.~\ref{fig:FSpara}(a). We will later choose a domain with $\bQ=(0.4\pi,0)$.
The possible case of a helical $2\bQ$- structure with a superposition containing both wave vectors 
will not be treated, it would require 4-component spinors in Hamiltonian below and therefore cause
an increase in algebraic complexity.}
The order parameter in Eq.~(\ref{eq:SDW}) breaks time 
reversal $\cal{T}$, inversion $\cal{I}$ and also reflection $(\sigma_v$, perpendicular to \bQ) and $C_4$
symmetries.
%which lifts the (pseudo-) spin Kramers degeneracy of the reconstructed conduction bands.
%For the commensurate AF case with $\bQ=(\pi,0), (0,\pi)$ or $(\pi,\pi)$ the product is however,
%still a discrete symmetry and thus the degeneracy of reconstructed bands is preserved.
The interaction of the localised  Gd spins $\bS(\bR_i)$ with conduction electron
 spins $\bs(\br)$ is assumed as an isotropic
contact exchange interaction $h_{ex}(\bR_i)=I_{ex}\bs(\br)\cdot\bS(\bR_i)\delta(\br-\bR_i)$,
which also leads to the effective RKKY interaction of Eq.~(\ref{eq:RKKY}).
The total Hamiltionian of coupled conduction and localised spins is described by the bilinear form
\cite{amici:00}
\bea
H=\sum_\bk(c^\dagger_{\bk\ua},c^\dagger_{\bk+\bQ\da})
\left(
 \begin{array}{cc}
\epsilon_\bk& \gamma\\
\gamma&\epsilon_{\bk+\bQ}\\
\end{array}
\right)
\left(
\begin{array}{c}
c_{\bk\ua}\\
c_{\bk+\bQ\da}\\
\end{array}
\right)
\label{eq:Ham}
\eea
with $\gamma=\frac{1}{2}I_{ex}S$ defining the interaction strength. 
It may be readily diagonalized by a Bogoliubov transformation
such that the reconstructed helical magnetic band states are mixtures of $|\bk\ua\ra$ and $|\bk+\bQ\da\ra$.
They are created by the rotated operators
\be
\bl
a^\dagger_{\bk +}&=
\cos\theta_\bk c^\dagger_{\bk\ua}-\sin\theta_\bk c^\dagger_{\bk+\bQ\da},
\\
a^\dagger_{\bk+\bQ -}&=
\sin\theta_\bk c^\dagger_{\bk\ua}+\cos\theta_\bk c^\dagger_{\bk+\bQ\da}.
\label{eq:recstate}
\el
\ee
Here the rotation angle is given by $\tan 2\theta_\bk=\ga/d_{\bk}$ with $\theta_\bk\in(-\frac{\pi}{4},\frac{\pi}{4})$. 
Furthermore we defined $d_\bk:=\fs(\eps_{\bk+\bQ}-\eps_\bk)$ which fulfils the relation $d_{\bk-\bQ}=-d_{-\bk}$. 
Explicitly we have
\bea
\left\{
\begin{array}{l}
\cos\theta_\bk\\
\sin\theta_\bk
\end{array}
\right.
=
\frac{1}{\sqrt{2}}\bigl[1\pm\frac{|d_\bk|}{(d_\bk^2+\ga^2)^\fs}\bigr]^\fs
\times
\left\{
\begin{array}{l}
1\\
s_\bk
\end{array}
\right.
\label{eq:recrot}
\eea
\\
where
 $s_\bk=
 {\rm sign}
 (\eps_{\bk+\bQ}-\eps_\bk)=
 {\rm sign} (d_\bk)$.
The mixing angle $\theta_\bk=\theta_{\bk\bQ}$ also depends on the fixed nesting vector
$\bQ$~but the latter index will be commonly suppressed. For the reconstructed helical bands
we use two different representations which we introduce now in detail for clarity in the subsequent development.
They correspond to two forms of the diagonalized  Hamiltonian given by
\be
\bl
H
&=
\sum_\bk[E_\bk^+a_{\bk +}^\dagger a_{\bk +} +E_\bk^-a_{\bk+\bQ -}^\dagger a_{\bk +\bQ-}]
\\
&=
\sum_{\bk}[\eps^+_\bk a_{\bk +}^\dagger a_{\bk +} +\eps^-_\bk a_{\bk -}^\dagger a_{\bk -}],
\el
\ee
where the second term in the first line has been shifted by $-\bQ$. This implies the identities
\bea
E^+_\bk=\eps^+_\bk ;\;\;\; E^-_{\bk}=\eps^-_{\bk+\bQ};\;\;\; \eps^-_\bk=E^-_{\bk-\bQ}.
\label{eq:helrel}
\eea
From the Bogoliubov transformation the $E^\pm_\bk$ are obtained as
\bea
E^\pm_{\bk}=\fs(\eps_{\bk+\bQ}+\eps_\bk)\mp s_\bk\bigl[\frac{1}{4}(\eps_{\bk+\bQ}-\eps_\bk)^2+\ga^2\bigr]^\fs.
\label{eq:recen}
\eea
Using the relations in Eq.~(\ref{eq:helrel}) and the inversion symmetry $\eps_\bk=\eps_{-\bk}$ of TB conduction bands
the proper helical reconstructed bands $\eps_\bk^\pm$ are given by 
\bea
\eps^\pm_\bk=\fs(\eps_{\bk\pm\bQ}+\eps_\bk)-s^\pm_\bk\bigl[\frac{1}{4}(\eps_{\bk\pm\bQ}-\eps_\bk)^2+\ga^2\bigr]^\fs,
\label{eq:helprop}
\eea
where  $s^\pm_\bk={\rm sign} (\eps_{\bk\pm\bQ}-\eps_\bk)$ and we abbreviated $s_\bk\equiv s^+_\bk$ before.
The helical band dispersions have no inversion symmetry which is broken by the order parameter in Eq.~(\ref{eq:SDW})
 but they fulfil the equivalent symmetry relations 
 \bea
 E^\pm_{-\bk}=E^\mp_{\bk-\bQ};\;\;\; {\text or}\;\;\;\eps_\bk^\pm=\eps^\mp_{-\bk}.
 \eea
 In the limit of vanishing exchange coupling $\gamma$ $\rightarrow$ $ 0$ we have $(E^+_\bk,E^-_\bk,)$ $\rightarrow$ $(\eps_\bk,\eps_{\bk+\bQ})$ and equivalently  $(\eps^+_\bk,\eps^-_\bk,)\rightarrow (\eps_\bk,\eps_\bk)$. Therefore the latter reduce to the twofold spin-degenerate conduction bands in this limit. In the perfect nesting case with commensurate AF order $\bQ=(\pi,\pi)$  one has $\epsilon_{\bk\pm\bQ}=-\epsilon_\bk$. Then the helical bands of Eq.~(\ref{eq:helprop}) simplify to $\eps^\pm_\bk=sign(\eps_\bk)(\eps_\bk^2+\gamma^2)^\fs$ which are degenerate throughout the BZ because the combination of  time reversal and lattice translation is a symmetry operation connected with a pseudospin-type degeneracy. Furthermore the dispersion has the AF gap $2\gamma$ on the equal energy surface $\epsilon_\bk=0$.
 
  While $\eps^\pm_\bk$ are the physical reconstructed bands in the following analytical treatment it is more convenient to use the related shifted dispersions $E_\bk^\pm$ for derivation of spectral densities and QPI quantities because they lead to more symmetric appearance of expressions. We remark that in any integral quantity involving the whole BZ both forms given in Eqs.~(\ref{eq:helprop},\ref{eq:recen}) may be used  leading to the same result since a shift by $\pm\bQ$ has no effect on the integrated quantity due to the periodicity. A comparison of their dispersion is shown in the example of Fig.~\ref{fig:dispplot}.
 \\
 
\begin{figure}
%\hspace{-0.5cm}
\includegraphics[width=1.05\columnwidth]{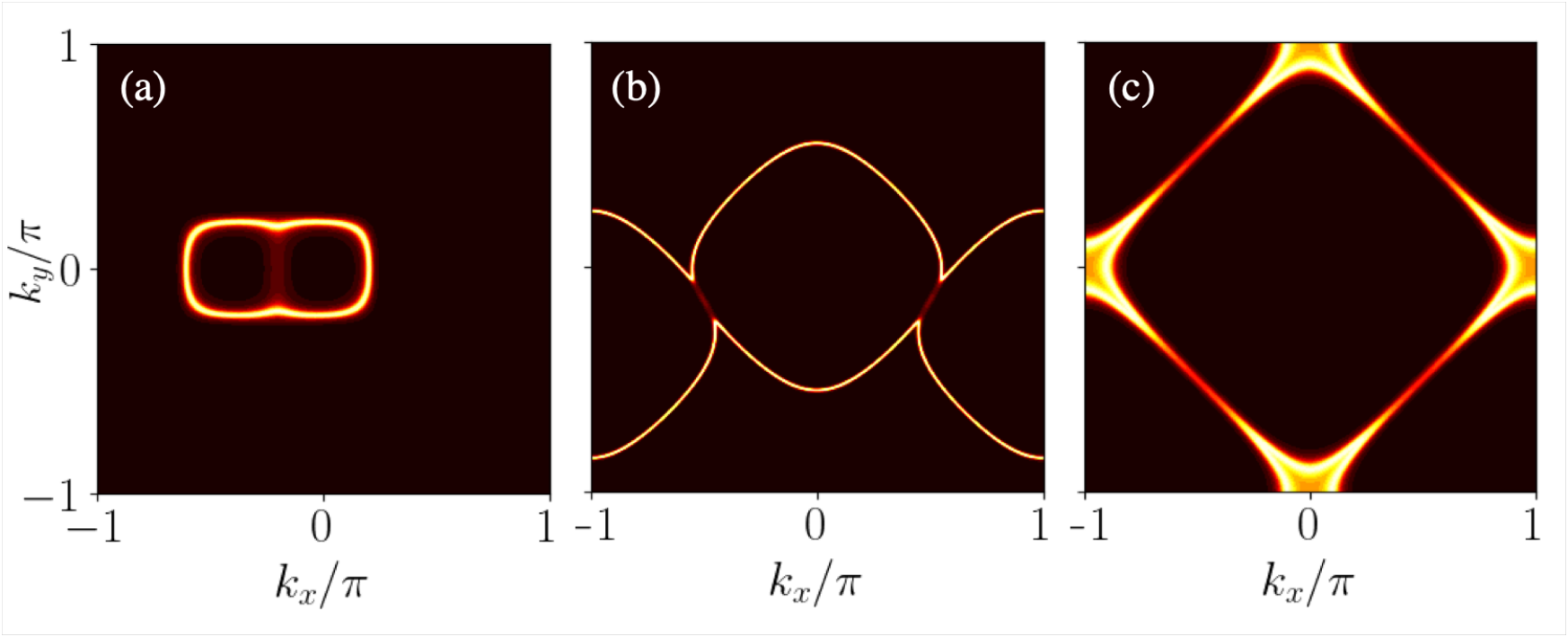}
\caption{Contour plot of sum of spectral functions of $R_{\ua}(\bk,\eps_F)+R_{\da}(\bk,\eps_F)=R_{+}(\bk,\eps_F)+R_{-}(\bk,\eps_F)$  (Eqs.~(\ref{eq:qpDOS},\ref{eq:regDOS})) of reconstructed  $E^\pm_{\bk}$  bands in the helical ordered state for interaction parameter $\gamma =0.15t$ with localised spins. The nesting vectors of bare bands are, consecutively $\bQ\simeq (0.4\pi,0), (\pi,0.3\pi), (\pi,\pi)$ as in Fig.~\ref{fig:FSpara}.  {\bf (a)} For nearly parabolic case IC nesting corresponding to $t=1.,t'=0.8,t''=-0.55$ and  $\eps_F/\eps_\Gamma=0.94$. {\bf (b)} intermediate case with IC nesting vectors  with  $t=1.,t'=0.7,t''=0.5$ and  $\eps_F/\eps_\Gamma=0.15$. {\bf (c)} Close to perfect commensurate nesting case with  $t=1.,t'=0.02,t''=0$ and  $\eps_F/\eps_\Gamma=0.025$. In this case inversion/reflection symmetries of the spectral functions is restored.}
\label{fig:specpara}
\end{figure}

\begin{figure}
%\vspace{0.3cm}
\includegraphics[width=0.99\columnwidth]{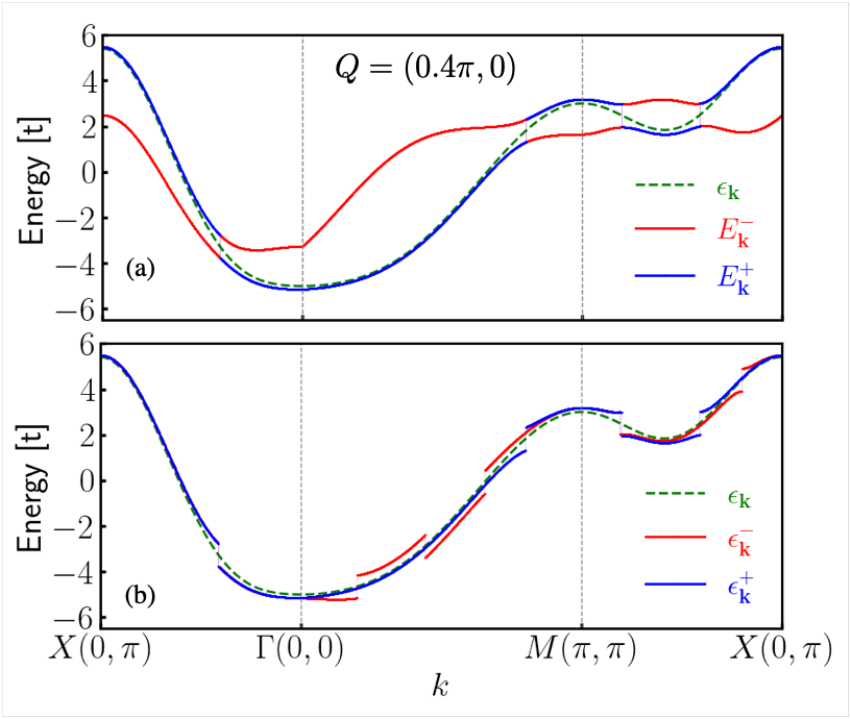}
\caption{
Reconstructed helical band dispersions $E^\pm_\bk$ (upper panel) and alternatively $\eps^\pm_\bk$ (lower panel). Here $\pm$ correspond to blue/red curves. Green-dashed line is the bare conduction band $\epsilon_\bk$ for TB parameters (Fig.~\ref{fig:FSpara}(a) with corresponding nesting vector $ \bQ=(0.4\pi,0.)$.
 Exchange coupling $\gamma=0.5t$ as in Fig.~\ref{fig:DOSplot} is applied. The splitting at the crossing points $\epsilon_\bk=\epsilon_{\bk+\bQ}$ is equal to $2\gamma$. Because  $\eps^\pm_\bk= \eps^\mp_{-\bk}$ their magnetic gaps appear at different \bk~vectors. They are degenerate along $X\Gamma$ line.
For $\gamma\rightarrow 0$ the correspondence is $(E^+_\bk, E^-_\bk)\rightarrow (\eps_\bk,\eps_{\bk+\bQ})$ and 
 $(\eps^+_\bk, \eps^-_\bk)\rightarrow (\eps_\bk,\eps_{\bk})$}
\label{fig:dispplot}
\end{figure}

\begin{figure}
%\vspace{0.5cm}
\includegraphics[width=0.99\columnwidth]{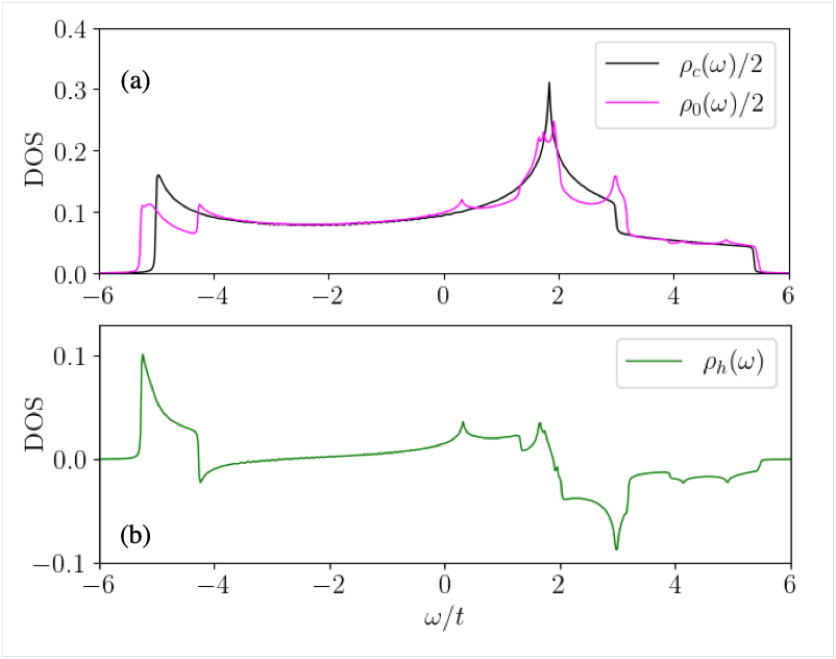}
%\vspace{-0.5cm}
\caption{Upper panel: DOS plot for TB conduction electron model $(\rho_c)$ with $t=1.,t'=0.8,t''=-0.55$ (black curve) corresponding to upper panels of Figs.~\ref{fig:FSpara}, to Fig.\ref{fig:specpara}(a) and to dashed green curve in Fig.~\ref{fig:dispplot}. The magenta curve ($\rho_0$) shows the quasiparticle DOS including the effect of coupling to the helical local moment structure with a coupling constant $\gamma/t=0.5$ and $\eta=0.01t$. It corresponds to the $\bq=0$ Fourier amplitude of the local DOS. Lower panel: Fourier amplitude of the satellite at $\bq=\pm\bQ$  corresponding to the spatially modulated DOS $\rho_h(\omega,\br)$ induced by the coupling to the helical local moment order.}
\label{fig:DOSplot}
\end{figure}

 Now we calculate the Green's function $G_0(\bk,\om)=(\om-h_\bk)^{-1}$ corresponding to the Hamiltonian in Eq.~(\ref{eq:Ham})
It is obtained as
\bea
\bl
G_0(\bk,\om)=&
\left(
\begin{array}{cc}
g_a(\bk,\om)& g_c(\bk,\om)\\
g_c(\bk,\om)&g_b(\bk,\om)
\end{array}
\right)\non\\[0.2cm]
=&
\frac{1}{D_\bk(\om)}
\left(
 \begin{array}{cc}
\om-\epsilon_{\bk+\bQ}& -\gamma
\\
-\gamma&\om-\epsilon_{\bk}\\
\end{array}
\right)
\label{eq:Green}
\el
\\
\eea
with the determinant given by
\bea
D_\bk(\om)&=&(\om-E_{\bk}^+)(\om-E_{\bk}^-)\non\\
&=&(\om-\epsilon_{\bk})(\om-\epsilon_{\bk+\bQ})-\gamma^2
.
\label{eq:deter}
\eea
The density of states (DOS) of the reconstructed bands is given by the integral over their associated spectral functions 
$R_n(\bk,\omega)$ $(n=\pm)$ shown in Fig.~\ref{fig:specpara}:
\bea
\hspace{-0.5cm}
\rho^0_n(\omega)=\frac{1}{N}\sum_\bk R_n(\bk,\omega); \;\; R_n(\bk,\omega)=\delta(\omega-E_{\bk}^n).
\label{eq:qpDOS}
\eea
The resulting total DOS is then  $\rho_0(\omega)=\sum_n\rho^0_n(\omega)$.
Due to the symmetry $E^n_{-\bk n}=E^{\bar{n}}_{\bk-\bQ};\; (\bar{n}:=-n)$ mentioned before they are equal for the two reconstructed bands, therefore $\rho^0_n(\omega)=\rho_0(\omega)/2$.
When the exchange coupling approaches zero, i.e. $\gamma\rightarrow 0$ this becomes
the bare conduction band DOS  $\rho_0(\omega)\rightarrow 2\sum_\bk\delta(\omega-\eps_\bk)=\rho_c(\omega)$. For finite $\gamma$ the splitting of reconstructed  $E^n_{\bk}$ bands (Fig.\ref{fig:dispplot})  leaves magnetic gap signatures in the total DOS of Fig~\ref{fig:DOSplot} (upper panel).\\ 
%
% %%%%%%%%%%%%%%%%%%%%% figure %%%%%%%%%%%%%%%%%%%%%%%%%%%%

%%%%%%%%%%%%%%%%%%%%%%fig%%%%%%%%%%%%%%%%%%%%%%%%%%%%%%%
%

Before we come to the main topic of STM spectrum and quasi particle interference it is necessary to discuss the  possible Fermi surface geometries of bare bands and the corresponding effective exchange interactions between the localised spins according to Eq.~(\ref{eq:sus}). We use parameters such that the conduction band bottom lies
at the $\Gamma$-point. This is shown in Fig.~\ref{fig:FSpara} for three typical cases
corresponding Fermi energy $\epsilon_F$ being close to band bottom $\epsilon_\Gamma$, for intermediate energies and for $\epsilon_F$ close to zero.\\
In Figs. 1(a) and 1(b)  parameters are chosen such that  the nesting vector $\bQ\sim (0.4\pi,0)$ parallel to $(100)$ approximately corresponds to that of the $\Gamma$-centered FS barrel in \GR~reported in Ref.~\onlinecite{bouaziz:22}. As shown in (b) the resulting effective interaction $J(\bq)\sim\chi(\bq)$ peaks around this wave vector which then leads to the ordering wave vector \bQ~of the localised moments $\bS_i$.\\
In Figs. 1(c) and 1(d) for somewhat larger $\epsilon_F$ the parameters are chosen such that
the nesting appears along two vectors $\bQ\simeq(\pi,0.3\pi)$ along off-symmetry direction and $\bQ'\simeq (0.7\pi,0.7\pi)$ along $(11)$ direction. The former is the dominant one as seen from the peak heights in Fig.1(d) which will prefer a helical local moment order with wave vector \bQ.\\
Finally in Figs. 1(e) and 1(f) we show the well known case of the almost perfect nesting situation with Fermi energy $\epsilon_F\simeq 0$ close to the band center. Then essentially only one strong peak at the commensurate position $\bQ=(\pi,\pi)$ survives in the effective interaction leading to the n.n. AF order of localised spins $\bS_i$.

\section{The theory of conduction electron STM spectrum with localised helical background magnetism}

{\BLU In the common scanning STM experiments the conductance $dI/dV(\br)$  is measured as function of position $\br$
of the nonmagnetic metallic tip on the sample surface. Here I(V) is the total (summed over spin directions) tunneling current and V the variable bias voltage between tip and sample. From the simple Bardeen formula for the tunneling current $I(V)$ the tunneling conductance may be approximately obtained as \cite{hoffman:11} as
\bea
\frac{dI}{dV}(\br)\approx -\frac{4\pi e^2}{\hbar}|M|^2\rho_{tip}(0)\rho(\omega=eV,\br)
\eea
where $\rho_{tip}(0)$ is the tip DOS and $|M|^2$ a squared tunneling matrix element, both assumed to be featureless
as function of energy $\omega$. Then the scan of the  tunnelling conductance is directly proportional to the 
position dependent conduction electron DOS $\rho(\omega,\br)$ at the surface. It has contributions from the 
background DOS of the lattice at the surface and from modulations coming from the scattering of conduction electrons
from impurities or defects. The latter contains important information on the conduction bands or rather their constant energy surfaces which is revealed in its momentum Fourier transform which is the QPI spectrum.
The influence of the ordered local moment structure on the itinerant electron tunneling conductance proportional to the  total surface DOS $\rho(\omega,\br)$ has two major consequences:}\\
Firstly, even in the pure case without any surface impurities the periodic surface
density of conduction electrons is modified due to their exchange interaction with the underlying magnetic
order of local moments. This interaction imprints the helical modulation of the latter with IC wave vector \bQ~also
on the reconstructed conduction electron states. Their corresponding density is then modified in such a way
that the Fourier transformed tunneling charge density exhibits satellite peaks at wave vectors $\pm\bQ$ relative to $\bk=0$, if we use plane waves for unreconstructed conduction states. If instead the cell periodic functions of inhomogeneous conduction electron background are used the charge-density satellites caused by the helical magnetic order will appear around all reciprocal wave vector $\bK_s=(2n\pi,2m\pi); \;s=(n,m)\in \cal{N}$, contained in the Fourier expansion of the periodic cell functions, i.e., at positions $\bK_s\pm\bQ$ (Appendix \ref{sec:app-cellper}). More general satellite positions are possible in the pure case and will be mentioned at the end of Sec.~\ref{sec:specpure}.\\
Secondly, if dilute impurities on the surface are present the quasiparticle interference originating from the impurity scattering leads to additional nonperiodic density modulations at all length scales. Their Fourier transform contains information on the reconstructed conduction band states. In particular concerning the gapping of these states close to the Fermi surface connected to the nesting properties with the spin density wave vector \bQ~and the breaking of inversion and reflection symmetry of conduction band states and reconstructed Fermi surface  due to the exchange interaction $\gamma$ with the helical spin structure. The QPI spectrum contains a regular part which survives when the coupling $\gamma\rightarrow 0$ and an anomalous helical part that vanishes in this limit.\\
We will consider both topographic spectral density of the pure system and QPI spectrum due to surface impurities  in the following within a unified theoretical framework based on the evaluation of the spatial representation of conduction electron Green's functions.
%
%
% %%%%%%%%%%%%%%%%%%%%% figure %%%%%%%%%%%%%%%%%%%%%%%%%%%%
%%%%%%%%%%%%%%%%%%%%%%fig%%%%%%%%%%%%%%%%%%%%%%%%%%%%%%%
%
% %%%%%%%%%%%%%%%%%%%%% figure %%%%%%%%%%%%%%%%%%%%%%%%%%%%
%%%%%%%%%%%%%%%%%%%%%%fig%%%%%%%%%%%%%%%%%%%%%%%%%%%%%%%
%

\subsection{The pure surface background spectrum from reconstructed bands}
\label{sec:specpure}

The local  DOS of the pure surface is obtained from the spatial  Green's function of the pure lattice as given by
\bea
G_0(\br,\br',\om)=\sum_\bk\Psi_\bk^\dag(\br')[\om-\hh_\bk]^{-1}\Psi_\bk(\br),
\label{eq:Greendef}
\eea
where the conduction states in spinor presentation are given by $\Psi^\dag(\br')=(\psi^*_{\bk\ua}(\br'),\psi^*_{\bk+\bQ\da}(\br'))$.
In plane wave approximation we have $\psi_{\bk\si}(\br)=(1/\sqrt{N})exp(i\bk\cdot\br)\chi_\si$ where $\chi_\si$ is the two-component 
spin wave function. The total pure surface (topographic) charge density connected with reconstructed conduction band states is then obtained as
\bea
\rho_t(\omega,\br)=-\frac{1}{\pi}\text{Im}G_0(\br,\br,\omega+i\eta).
\eea
By replacing the wave functions $\Psi_\bk(\br)\rightarrow \fs[1\pm\si_z] \Psi_\bk(\br)=\Psi_\bk^{\ua\da}(\br)$ with the spin projected ones $(\si=\ua,\da)$ we may also derive the spin-projected densities $\rho^0_\si(\omega,\br)$ from the corresponding Green's functions  $G^0_\si(\br,\br,\omega)$ by the equivalent formula. Expressed in terms of the Green's- function matrix elements of Eq.(\ref{eq:Green}) we obtain the spatially homogeneous spin projected Green's function
\bea
G^0_{\ua,\da}(\omega)=
\frac{1}{N}\sum_\bk g_{a,b}(\bk,\omega),
%;\ G^0_\da(\omega)=\frac{1}{N}\sum_\bk g_b(\bk,\omega);
\eea
and the spatially dependent total Green's function 
\bea
\bl
&G_0(\br,\br,\omega)
=
\\
&
\frac{1}{N}\sum_\bk\{ 
[g_a(\bk,\omega)
\!+\!
g_b(\bk,\omega)]
+g_c(\bk,\omega)(e^{i\bQ\cdot\br}
\!+\!
e^{-i\bQ\cdot\br})\}.
\non
\label{eq:purelocal}
\el
\\
\eea
Note that this expression contains an additional term to the sum of the spin projected Green's function because the $\ua,\da$ spin states are not eigenstates due to the presence of the coupling to the helical local moment order. From this we obtain the reconstructed and spin projected conduction electron spectral functions and densities as
\bea
\label{eq:recdos}
\bl
R_{\ua\da}(\bk,\omega)&=
\cos^2\theta_\bk\delta(\omega-E^\pm_{\bk})+\sin^2\theta_\bk\delta(\omega-E^\mp_{\bk});
\;
\\
\rho^0_{\ua\da}(\omega)
&=
\frac{1}{N}\sum_\bk R_{\ua\da}(\bk,\omega)
\label{eq:regDOS}
\el
\eea
We note that the sum of the spin projected spectral functions $R_\ua +R_\da=R_+ + R_-$ is equal to the sum of the
quasiparticle specral functions in Eq.~(\ref{eq:qpDOS}).
Although superficially the above $\rho^0_{\ua\da}$ densities look different compared to $\rho^0_\pm$ of Eq.~(\ref{eq:qpDOS}), using again the symmetry relation $E^n_{-\bk}=E^{\bar{n}}_{\bk-\bQ};\; (\bar{n}:=-n)$,
they turn out to be identical, i.e. $\rho^0_{\ua\da}=\rho^0_\pm\equiv \rho_0/2$ and are given
by a single DOS function $\rho_0(\omega)$ shown in Fig.~\ref{fig:DOSplot} (blue curve).\\
 
The total reconstructed spectral charge density in real space is given by
\bea
\bl
\label{eq:totDOS}
&\rho_t(\omega,\br)=\rho_0(\omega)+\rho_{h}(\omega,\br)
\\
&=\frac{1}{N}\sum_{\bk, n=\pm}\delta(\omega-E^n_{\bk})\\
&+\frac{1}{N}\sum_{\bk, n=\pm} n\sin\theta_{\bk}\cos\theta_{\bk}\delta(\omega-E^n_{\bk})
(e^{i\bQ\cdot\br}
\!+\!
e^{-i\bQ\cdot\br}).
\el
\label{totDOS}
\eea
Here the first part is simply the  homogeneous quasiparticle density of Eq.~(\ref{eq:qpDOS}) or Eq.~(\ref{eq:recdos}) given by $\rho_0(\omega)$ and the second is a spatially modulated contribution. Similar to Eq.~(\ref{eq:regDOS}) its amplitude $\rho_h(\omega)$ is obtained from the helical spectral function $R_h(\bk,\omega)$ according to
\bea
\bl
R_h(\bk,\omega)&=
\sum_{n=\pm}R^h_n(\bk,\omega)
\\
&=
\sum_{n=\pm} n\sin\theta_{\bk}\cos\theta_{\bk}\delta(\omega-E^n_{\bk}),
\\
\rho_h(\omega)&=
\sum_{n=\pm}\rho^h_n(\omega)=
\frac{1}{N}\sum_{\bk,n} R^h_n(\bk,\omega).
\label{eq:helDOS}
\el
\eea
Its appearance is a direct consequence of the helical ordered local moments background. For vanishing exchange coupling to the local moments this term also vanishes since $\sin\theta_\bk\sim \gamma$. For finite $\gamma$ it fulfils the sum rule $\int\rho_h(\omega)d\omega =0$ {\BLU since the total homogeneous charge density is unchanged by the exchange scattering}.  Again due to the symmetries mentioned before we have $\rho^h_\pm(\omega)=\rho_h(\omega)/2$.
Taking the momentum Fourier transform we obtain
\bea
\rho_t(\omega,\bq)=\rho_0(\omega)\delta_{\bq, 0}+\rho_h(\omega)[\delta_{\bq +\bQ}+\delta_{\bq -\bQ}],
\label{eq:totDOS1}
\eea
where the first part appears at zero momentum position and the second one at the helical satellite positions $\pm\bQ$. The oscillating part $\rho_h(\omega,\br)$ changes sign as function of position as well as frequency,  the total DOS,  $\rho_t(\omega,\br)$,  is always positive as required physically and shown by the following argument:
Consider the maxima/minima of  $\rho_h(\omega,\br)$  that appear at real space positions given by $\cos(\bQ\cdot\br)=\pm 1$. 
At these positions we have, according to Eq.~(\ref{eq:totDOS}) $\rho^\pm_t(\omega)=\rho_0(\omega)\pm\rho_h(\omega)$. Then the total density may be transformed, using also Eq.~(\ref{eq:recdos}) into
\bea
\rho^\pm_t(\omega)=\frac{1}{N}\sum_{\bk,n}(\cos\theta_\bk\pm n\sin\theta_\bk)^2\delta(\omega-E^n_{\bk}),
\eea
which is evidently positive. Since the modulus of the cosine is smaller for any other position, $\rho_h(\omega,\br)$ will be positive throughout. For \bQ~$\parallel$$(1,0)$ the extremal positions are $x_m^+=m\lambda_h$ and $x_m^-=(m+\fs)\lambda_h$, respectively, where $\lambda_h=2\pi/Q$ is the wavelength of the helical spin structure. These are also the spatial maxima/minima for the total density  $\rho_h(\omega,\br)$ if  $\rho_h(\omega)>0$  and vice versa for  $\rho_h(\omega)<0$.\\
Finally we note again that if we do not restrict to plane wave conduction states but use general 
cell-periodic Bloch functions as starting point (Appendix \ref{sec:app-cellper}) the density peaks of $\rho(\omega,\bq)$ appear at positions $\bK_s$ and $\bK_s\pm\bQ$, respectively where $\bK_s$ is a reciprocal lattice vector (including $\bK_s=0$ of the plane wave case). Since we used a single-\bQ~helix local moment order only $\pm\bQ$ satellites will appear. It is reasonable to 
expect that for a double-\bQ~structure with $\bQ_{1,2} \parallel a,b$ both $\bQ_{1,2}$ satellites are present. The harmonic form of Eq.~(\ref{eq:SDW}) will not support, however, satellites at $2\bQ$ positions as have been observed \cite{spethmann:24}. For this purpose one presumably needs to include higher harmonics in the helix structure
of Eq.~(\ref{eq:SDW}). 

The calculated Fourier transform of the pure surface density according to Eq.~(\ref{eq:totDOS1}) is shown in the contour plot of Fig.~\ref{fig:satellites}. It exhibits the main central peak of the homogeneous background density and the two satellites at $\pm\bQ$ due to the influence of the helical local moment order and their relative amplitude depends on $\omega$.

{\BLU We have shown before in Eq.(\ref{eq:regDOS}) that the helical rotating structure of local moments does not lead
to a uniform spin polarization of the conduction electron densities $\rho_{\ua\da}(\omega)$ caused by exchange coupling.} However, It is natural to expect that the localised spin helix of Eq.~(\ref{eq:SDW}) will also induce a polarized conduction electron  helical spin structure $\la\bs(\bR_i)\ra$ via the contact exchange interaction $h_{ex}$ given again by the form
\bea
\la\bs(\bR_i)\ra=s(T)({\bf a}\cos\bQ\cdot\bR_i+{\bf b}\sin\bQ\cdot\bR_i).
\label{eq:condpol}
\eea
The temperature dependent induced  amplitude $s(T)$ of helical conduction electron spin structure may be obtained by using the reconstructed band states in Eq.~(\ref{eq:recstate}) for calculating the expectation values. This leads to the result
\bea
s_0(T)=\fs\sum_\bk\sin\theta_\bk\cos\theta_\bk[f(E^-_\bk)-f(E^+_\bk)],
\label{eq:polamp}
\eea
where $f(E)$ denotes the Fermi function. Here we assumed that
the background localised helix amplitude is at saturation value $S$ in the relevant temperature range.
The spin polarization increases monotonically with exchange coupling $\gamma$, with the strongest increase
observed for the perfect nesting case. {\BLU This dependence is shown in Fig.~\ref{fig:satellites}(b) for the three FS models.}

\subsection{The QPI spectrum 
%of G\lowercase{d}R\lowercase{u}$_2$S\lowercase{i}$_2$
for scalar and exchange scattering}
\label{sec:QPI}

\subsubsection{A model for scalar and exchange impurity scattering}
\label{sec:scpot}

The top layer in the experiments of Ref.~\onlinecite{spethmann:24} was a Gd- layer containing the periodic
arrangement of S=7/2 Gd$^{3+}$ spins which result from strongly bound 4f- electrons that do not directly contribute
to the tunneling current. Their only effect in the STM context is the reconstruction of conduction electrons states.  Assuming that they are frozen at low temperatures (neglecting magnons) they do not lead to any further scattering of the already reconstructed band states. Their scattering can however be achieved by impurities on the surface causing additional density modulation of  reconstructed conduction electron states at all wavelengths seen in the QPI image. We consider a scalar $(V_s)$ as well as an exchange $(V_{ex})$ impurity scattering term.
The former is due to nonmagnetic impurities or defects while the latter is due to extra magnetic ions on interstitial sites of the surface that generally also have a spin $S_{imp}$ different from the spin $S$ of the periodic magnetic background order constructed from the Gd ions. For simplicity we assume that the impurity spin is of Ising type is oriented perpendicular to the surface by the local crystalline electric field providing a static exchange scattering potential. 
Together the scattering potential at the impurity site reads
\bea
\hspace{-0.5cm}
\hat{V}(\bq)=V_s(\bq)\sigma_0+V_{ex}(\bq)\sigma_z=
{\rm diag}
\Big[
V_+(\bq),V_-(\bq)
\Big],
\label{eq:Vperp}
\eea
where $\si_0,\si_z$ are unit and Pauli matrix in the space of conduction electron spin. In the following we
assume a contact interaction with the impurity leading to a $\hat{V}$ independent of momentum transfer \bq , the latter will therefore generally be suppressed. The diagonal elements are then given by $V_{\pm}=V_s\pm V_{ex}$, respectively.
{\BLU The scalar scattering may be caused by adatoms, interstitials or vacancies on the surface, which will modify or cut nearest neighbor hopping bonds of conduction electrons. Then one may expect that the scattering potential $V_s$ of periodic Bloch states from the scalar defect site will be of similar order of magnitude as the hopping energy $t$.}
%
% %%%%%%%%%%%%%%%%%%%%% figure %%%%%%%%%%%%%%%%%%%%%%%%%%%%
\begin{figure}
%\vspace{0.5cm}
%\includegraphics[width=0.70\columnwidth]{satellites}
%\includegraphics[width=0.70\columnwidth]{spinpol}
\includegraphics[width=1.0\columnwidth]{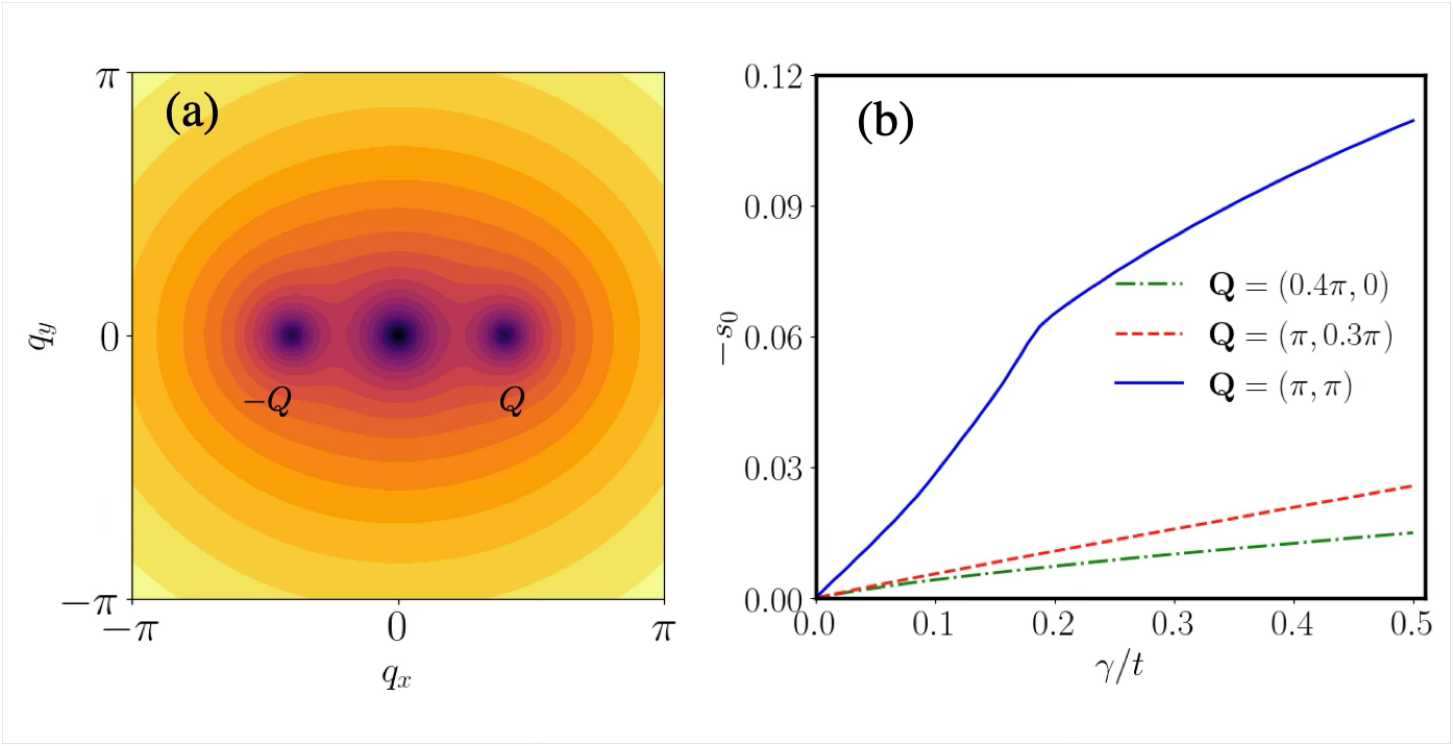}
\caption{a) Background pure surface spectrum of Eq.~(\ref{eq:totDOS1}) showing the central peak and the helical satellites at $\pm\bQ$ for the case of Fig.~\ref{fig:FSpara} (a,b). The amplitude $\rho_t(\omega,\bq)$ at $\omega/t=-5.1$  is shown in a logarithmic plot with assumed finite momentum width $\delta q/\pi=0.005$ in the BZ.
{\BLU b) Helical spin polarization as function of exchange interaction for the three FS models of Fig.~\ref{fig:FSpara}
according to Eq.~(\ref{eq:polamp}) at T=0.}}
\label{fig:satellites}
\end{figure}
%%%%%%%%%%%%%%%%%%%%%%fig%%%%%%%%%%%%%%%%%%%%%%%%%%%%%%%
%
%
% %%%%%%%%%%%%%%%%%%%%% figure %%%%%%%%%%%%%%%%%%%%%%%%%%%%
\begin{figure*}
\includegraphics[width=0.82\textwidth]{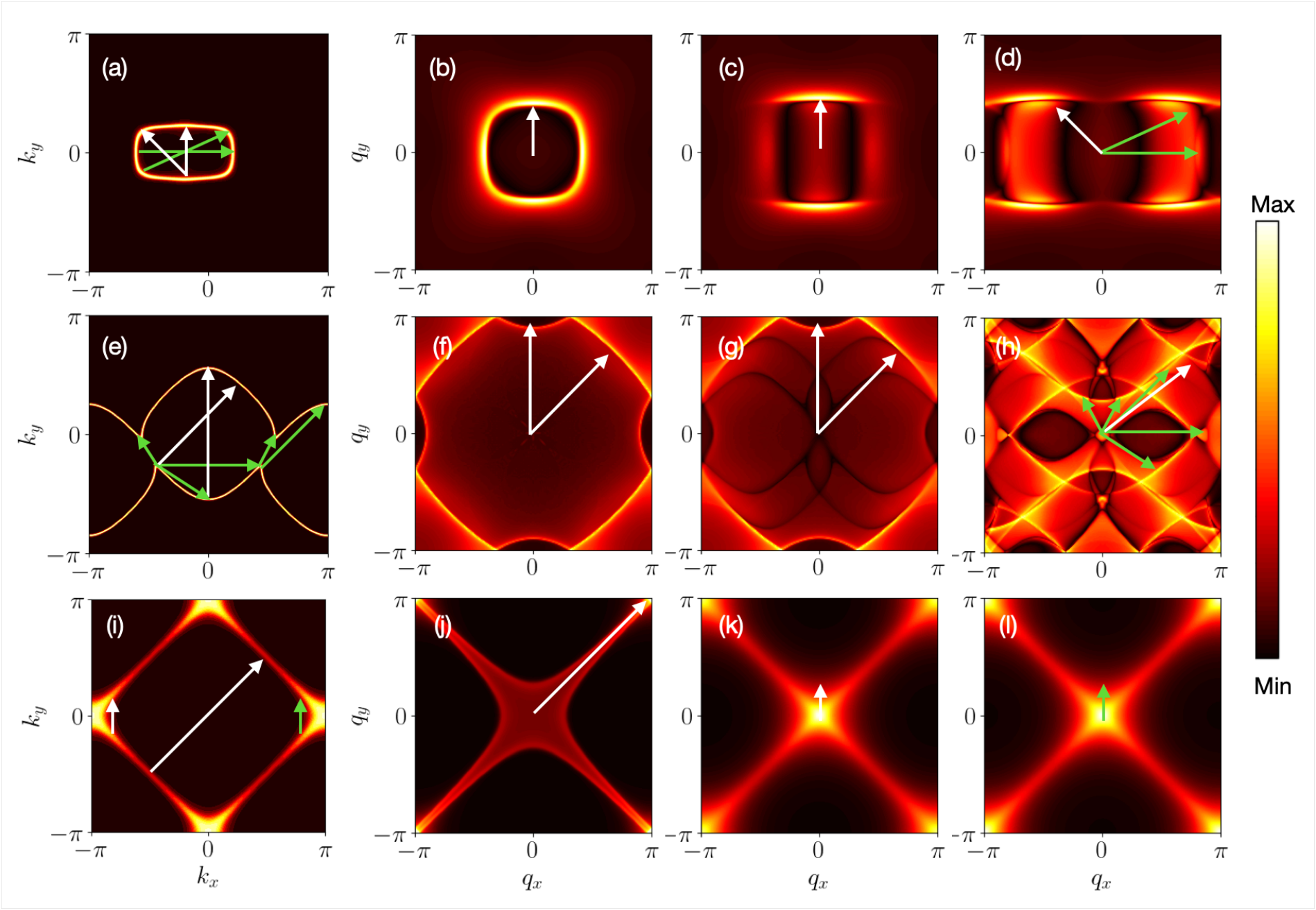}
\caption{ QPI spectra according to Eq.~(\ref{eq:delrho}). The rows correspond
to the three Fermi surface models with their associated \bQ~vectors (Fig.~\ref{fig:FSpara}), left column).
In each case $\hat{\omega}=0$ where $\hat{\omega}=eV= \omega-\eps_F$ (V= bias voltage)
is measured from the Fermi energy of each model.
The leftmost column shows again the spectral functions of Fig.~\ref{fig:specpara} $(\gamma/t=0.25)$ with indication of characteristic scattering wave vectors without (white) and with (green) involving the chosen helical wave vector $\pm\bQ$.
As a reference the second column is the QPI image without the helical order $(\gamma =0)$ with only the former type of momentum transfer appearing. It maps the simple Fermi surface with a doubling of the radius.
 The third and last column show the regular $\delta\rho_0$ and anomalous 
helical $\delta\rho_h$ contribution to the QPI image for finite $\gamma/t=0.25$, respectively (absolute values are plotted)
The former is related to momentum transfer from impurity scattering only (white arrows), the latter involves in addition the ordering vector leading to $\bq\pm\bQ$ total momentum transfer. In the lower row perfect nesting FS with commensurate $\bQ=(\pi,\pi)$ they may be partly the same.
{\BLU In particular the model in the first row shows strong $C_4$ symmetry breaking associated with the domain choice of  $\pm\bQ=\pm(0.4\pi,0)$.
For the choice  $\pm\bQ=\pm(0,0.4\pi)$ its QPI patterns would be rotated by $\pi/2$ (see also Figs.~\ref{fig:QPI_omvar},\ref{fig:QPI_gavar}).}
}
\label{fig:QPI_om0}
\end{figure*}
%%%%%%%%%%%%%%%%%%%%%%fig%%%%%%%%%%%%%%%%%%%%%%%%%%%%%%%
%
\subsubsection{The local DOS and its momentum Fourier transform}
\label{sec:Born}

In addition to the topographic periodic density of the pure surface treated in Sec.\ref{sec:specpure}  the scattering
from random impurities at the surface leads to further contributions $\delta\rho(\omega,\br)$ {\BLU such that the total spatially modulated
DOS is then $\rho(\omega,\br)=\rho_t(\omega,\br)+\delta\rho(\omega,\br)$.} The impurity part  $\delta\rho(\omega,\br)$ is  nonperiodic, i.e. contributes  at all wave vectors \bq~of the BZ and originate from the quasiparticle interference \cite{capriotti:03,farrell:15} of single or repeated scattering of conduction electrons at a single impurity site. The QPI density contribution may be evaluated from the impurity correction to the real space Green's function matrix according to $\delta \rho(\br)=(-1/\pi) Im \delta G(\br,\br)$ where the correction is given by (frequency $\om$ suppressed)
\bea
\bl
\delta G(\br,\br')&=
G(\br,\br')-G_0(\br,\br')
\\
&=\sum_{\bk\bq}\Psi^\dag_{\bk-\bq}(\br')\delta G(\bk,\bk-\bq)\Psi_\bk(\br).
\el
\eea
As long as the impurity potential is sufficiently weak to avoid local bound state formation the  scattering may be treated in Born approximation. {\BLU For strong scattering potential the description of bound states requires the full T matrix approach (Sec.~\ref{sec:tmat}). We mentioned that a contact scattering on the impurity site leads to a momentum-independent scattering potential. Then the QPI pattern in momentum space and frequency dependence of T matrix that contains the bound state pole are essentially decoupled \cite{akbari:13}. Therefore for the purpose of studying
the influence of helical local moments on the momentum structure of the QPI spectrum we may restrict to a weak impurity scattering situation in Born approximation. In the two limits we consider a single scattering event (Born case) or repeated
scatterings (T matrix case) from a {\it single} impurity site adequate for the dilute limit. Scattering from multiple impurity sites
for larger concentrations has been considered in Ref.~\cite{capriotti:03} and was found to preserve the essential QPI momentum structure but  lead to a degradation of its resolution.

Then one obtains for the Fourier transform of the scattering induced correction to the Green's function in Born approximation:}

\bea
\bl
\hspace{-0.5cm}\delta G(\bk,\bk-\bq,\om)
&=G_0(\bk,\om)\hat{V} G_0(\bk-\bq,\om)
\\
&=[\om-\hh_\bk]^{-1}\hat{V}[\om-\hh_{\bk-\bq}]^{-1},
\label{eq:Greencorr1}
\el
\eea
which may be written explicitly in matrix form, similar to Eq.~(\ref{eq:Green}) as
\be
\bl
\delta G(\bk,\bk-\bq,\om)
&=\left(
\begin{array}{cc}
\delta g_a(\bk,\bq,\om)& \delta g_c(\bk,\bq,\om)\\
\delta g'_c(\bk,\bq,\om)&\delta g_b(\bk,\bq,\om)
\end{array}
\right),
\el
\label{eq:Greencorr2}
\ee
where the individual matrix elements are obtained as 
\bea
\delta g_\alpha=
\frac{
\hat{\delta}g_\alpha
}{
(D_\bk D_{\bk-\bq})}
;\
\quad
(\alpha =a,b,c) 
,
\label{eq:gfunc}
\eea
with the denominator from Eq.~(\ref{eq:deter}) and the numerators
\bea
\bl
\hde g_a&=\;\;V_+(\om-\eps_{\bk+\bQ})(\om-\eps_{\bk-\bq+\bQ})+V_-\ga^2
\\
\hde g_b&=\;\;V_-(\om-\eps_{\bk})(\om-\eps_{\bk-\bq})+V_+\ga^2
\\
\hde g_c&=-\ga[V_+(\om-\eps_{\bk+\bQ})+V_-(\om-\eps_{\bk-\bq})]
\\
\hde g'_c&=-\ga[V_+(\om-\eps_{\bk-\bq+\bQ})+V_-(\om-\eps_{\bk})].
\label{eq:hgfunc}
\non
\el
\\
\eea
Then in equivalence to Eq.~(\ref{eq:purelocal}) for the pure surface we obtain
the impurity correction to the local Green's function
\bea
\bl
\label{eq:implocal}
\delta G(\br,\br,\om)
=&\frac{1}{N}\sum_{\bk\bq} e^{i\bq\br}\{ [\delta g_a(\bk,\bq)+\delta g_b(\bk,\bq)\\
&+
\delta g_c(\bk,\bq)e^{i\bQ\cdot\br}+\delta g'_c(\bk,\bq)e^{-i\bQ\cdot\br})\}_{\om}.
\non
\el
\\
\eea
Finally the Fourier transform of the QPI spectral density is obtained as
\bea
\delta\rho(\omega,\bq)=-\frac{1}{\pi}\text{Im}\sum_\br\delta G(\br,\br,\omega+i\eta)e^{i\bq\br}.
\label{eq:impspec1}
\eea
Using Eqs.~(\ref{eq:implocal},\ref{eq:impspec1}) this may be expressed as 
%$(\om\rightarrow \omega +i\eta)$
%
\bea
\bl
\delta\rho(\omega,\bq)
&=\delta\rho_0(\omega,\bq)+\delta\rho_{h}(\omega,\bq),
\\
\delta\rho_0(\omega,\bq)
&=
\frac{-1}{\pi N}\text{Im}
\sum_\bk[\de g_a(\bk,\bq)+\de g_b(\bk,\bq)]_{\omega +i\eta},
\\
\delta\rho_h(\omega,\bq)
&=
\\
&\hspace{-1cm}
\frac{-1}{\pi N}\text{Im}
\sum_\bk[\de g_c(\bk,\bq-\bQ)+\de g'_c(\bk,\bq+\bQ)]_{\omega +i\eta}.
\non
\label{eq:rhofunc}
\el
\\
\eea
The density is composed of two parts: A regular one $(\de \rho_0)$ where the 
momentum transfer \bq~is only coming from the impurity scattering  and a second anomalous helical
part  $(\de \rho_h)$ where the momentum transfer $\bq\pm\bQ$ is coming from impurity scattering
as well as helical modulated magnetic background. In other words for every regular density contribution
at wave vector $\bq$ one has two helical satellite contributions at $\bq\pm\bQ$ due to the magnetic order, similar to the situation for the pure surface density.\\

The two spectral density contributions $\de\rho_0$ and $\de\rho_h$
may be explicitly evaluated  by using Eqs.~(\ref{eq:gfunc},\ref{eq:hgfunc},\ref{eq:rhofunc}) and then we get for the regular and anomalous helical contributions to the total QPI spectrum $\delta\rho=\delta\rho_0+\delta\rho_h$ $(\om\rightarrow \omega +i\eta)$:

\begin{widetext}
\bea
\bl
\delta\rho_0(\omega,\bq)
&=
-\frac{1}{\pi N}\text{Im}\sum_{\bk}\Bigl[
\frac{V_+(\om-\eps_{\bk+\bQ})(\om-\eps_{\bk-\bq+\bQ}) + 
V_-(\om-\eps_{\bk})(\om-\eps_{\bk-\bq})+\gamma^2(V_+ +V_-)} 
{(\om-E^+_{\bk})(\om-E^-_{\bk})(\om-E^+_{\bk-\bq})(\om-E^-_{\bk-\bq})}
\Bigr]
\\[0.2cm]
\delta\rho_h(\omega,\bq)
&=
\;\;\;\frac{\gamma}{\pi N}\text{Im}\sum_{\bk}\Bigl[
\frac{V_+(\om-\eps_{\bk+\bQ})+V_-(\om-\eps_{\bk-\bq+\bQ})}
{(\om-E^+_{\bk})(\om-E^-_{\bk})(\om-E^+_{\bk-\bq+\bQ})(\om-E^-_{\bk-\bq+\bQ})}
+
\\
&
\hspace{2cm}\frac{V_+(\om-\eps_{\bk-\bq})+V_-(\om-\eps_{\bk})} 
{(\om-E^+_{\bk})(\om-E^-_{\bk})(\om-E^+_{\bk-\bq-\bQ})(\om-E^-_{\bk-\bq-\bQ})}
\Bigr].
\label{eq:delrho}
\el
\eea
\end{widetext}
There is an important distinction between the two contributions. The regular part
$\de\rho_0$ survives in the limit $\gamma\rightarrow 0$ when the coupling to helical magnetic order vanishes
and it simply turns into the QPI spectrum of the unreconstructed conduction electrons. This may be explicitly demonstrated
by noting that $E^+_\bk\rightarrow \epsilon_\bk$ and $E^-_\bk\rightarrow \epsilon_{\bk+\bQ}$ in that limit. Then, inserting in the above expression for $\de\rho_0$ we get
\bea
\de\rho_0(\omega,\bq)=-\frac{2}{\pi N}\text{Im}\sum_\bk\frac{V_+ +V_-}{(\om-\eps_\bk)(\om-\eps_{\bk-\bq})},
\eea
which is the regular QPI spectrum for the unreconstructed bare conduction bands (the factor $2$ is due to the spin degeneracy, furthermore $V_++V_-=2V_s$).
On the other hand all terms
in the anomalous  helical part $\delta_h$, induced by the satellite scattering are proportional to $\gamma V_\pm$, thus they require the presence of  helical order simultaneously with the presence of surface impurities. Therefore for unreconstructed conduction electrons with vanishing $\gamma$ the anomalous helical density $\delta_h$ is absent. This is akin to the situation for the pure surface where the satellites at $\pm \bQ$ positions in the BZ also vanish in this limit.
We finally note that $\delta\rho_{0,h}(\omega,\bq)\neq \delta\rho_{0,h}(\omega,-\bq)$ generally and therefore, like the spectral functions in Fig.~\ref{fig:specpara} are not inversion or reflection symmetric. However, since $E^\pm_\bk(\bQ)=E^\pm_{-\bk}(-\bQ)$ (Eq.~(\ref{eq:recen})) we have the symmetry $\delta\rho_{0,h}(\omega,\bq)_\bQ=\delta\rho_{0,h}(\omega,-\bq)_{-\bQ}$ under simultaneous reversal of all momenta.
The helical state with momentum $-\bQ$ is simply a phase shifted form of the modulation in Eq.~(\ref{eq:SDW}).

\section{T matrix calculation of impurity bound states in the helical background}
\label{sec:tmat}

In the previous sections we first discussed the influence of helical background magnetic order on the pure surface topographical
spectrum, demonstrating the appearance of the satellite peaks and secondly we investigated to which extent the 
conduction bands exhibit the helical order reconstruction features in the momentum dependent QPI spectrum resulting from scattering due to surface impurities.
Thereby we used the Born approximation for weak impurity scattering since the momentum structure of the
QPI image is little affected by the higher order processes provided the scattering potential $\hat{V}$ is independent of momentum
transfer \bq. On the other hand if the impurity scattering potential is strong enough a local bound or state at the impurity sites may form below the conduction electron continuum \cite{balatsky:06}  and it is expected that they too may be markedly affected if a strong helical magnetic order of local moments exists that interferes with the bound state formation via the contact exchange interaction. This topic will be investigated in the present section.\\

The local spectrum at the impurity site $\br=0$ beyond Born approximation may also be obtained from the Green's function correction $\delta G(\br,\br,\om)$ but replacing the bare scattering potential $\hat{V}$ by the full T matrix. In analogy to Eq.~(\ref{eq:Greencorr1}) we then get
\bea
\bl
\hspace{-0.3cm}\delta G(\om)
&=
\sum_{\bk,\bq}G_0(\bk,\om)T(\om) G_0(\bk-\bq,\om)
\\
&=
\sum_{\bk,\bk'}[\om-\hh_\bk]^{-1}T(\om)[\om-\hh_{\bk'}]^{-1},
\label{eq:Greencorr3}
\el
\eea
with $\bk'=\bk-\bq$. Here the T matrix is evaluated summing repeated single impurity scattering processes \cite{akbari:13,farrell:15,kaasbjerg:20} which leads to
\bea
\bl
T(\om)
=&[1-\hV G_0(\om)]^{-1}\hV
,\\
G_0(\om)
=&\frac{1}{N}\sum_\bk G_0(\bk,\om)=
\left(
\begin{array}{cc}
g_d(\om)&g_c(\om)\\
g_c(\om)&g_d(\om)
\end{array}
\right)\non
\label{eq:Tmat1}
\el
\eea
with the matrix elements given by
\bea
g_d(\om)=
%\frac{1}{N}\sum_\bk\frac{\om-\eps_\bk}{(\om-E_\bk^+)(\om-E_\bk^-)}=
\int\frac{\fs\rho_0(\omega')}{\om-\omega'}d\omega';\;\;
%\non\\
g_c(\om)=
%\frac{1}{N}\sum_\bk\frac{(-\ga)}{(\om-E_\bk^+)(\om-E_\bk^-)}=
\int\frac{\rho_h(\omega')}{\om-\omega'}d\omega'
.
\non\\
\label{eq:Greenint}
\eea
%
% %%%%%%%%%%%%%%%%%%%%% figure %%%%%%%%%%%%%%%%%%%%%%%%%%%%
\begin{figure}
%\hspace{-0.5cm}
\includegraphics[width=1.0\columnwidth]{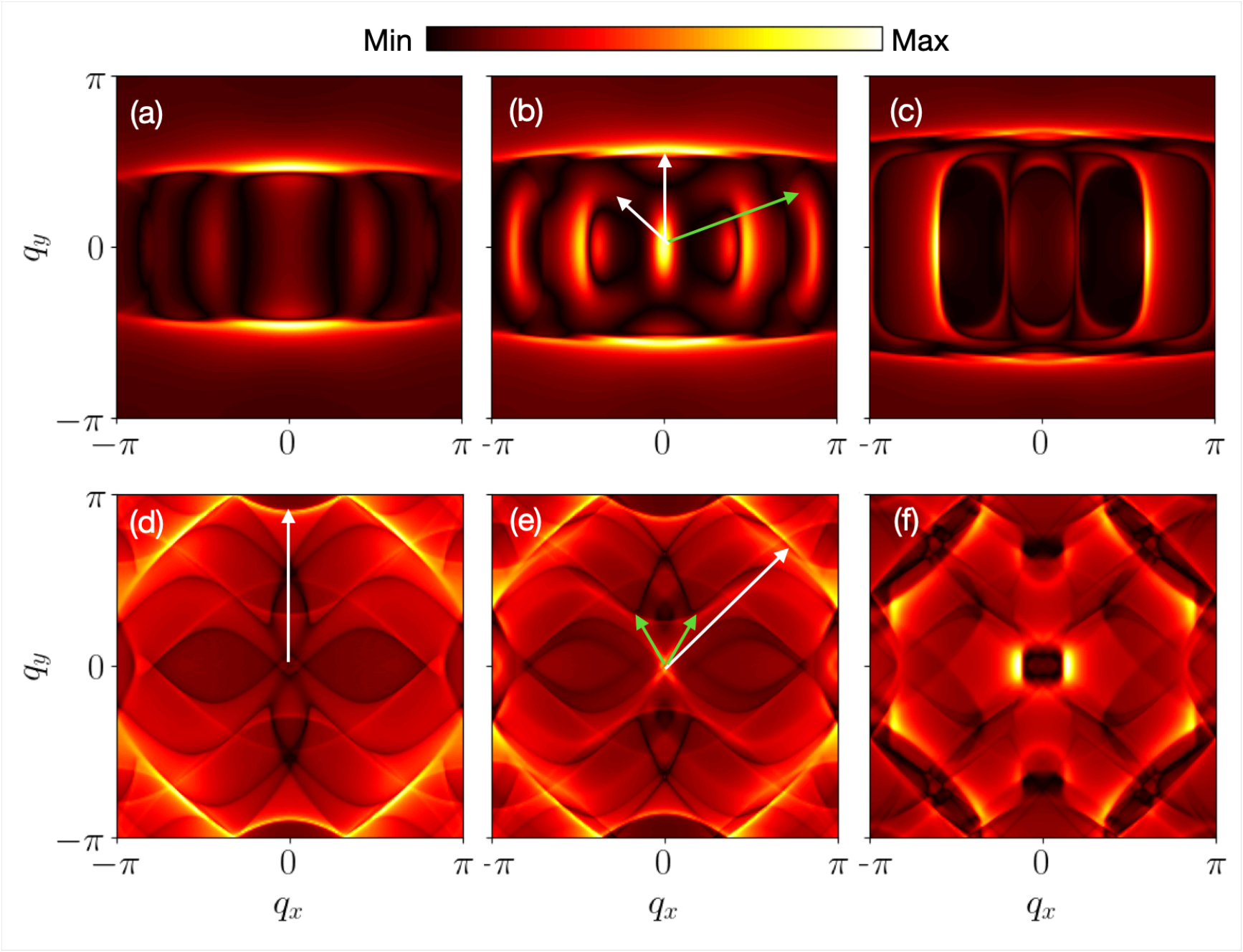}
\caption{ Total QPI spectra $\delta\rho_t(\omega,\bq)=\delta_0+\delta_h$ (modulus) according to Eq.~(\ref{eq:delrho}) for $\gamma=0.25$. 
The two rows correspond to the first two Fermi surface models with their associated ordering \bQ~vectors in Fig.~\ref{fig:FSpara} (left column). The three columns in this figure are for frequencies $\hat{\omega}=0., 0.11, 0.28.$ For moderate frequency the 
similar scattering contributions as explained in Fig.~\ref{fig:QPI_om0} may be identified.}
\label{fig:QPI_omvar}
\end{figure}
%%%%%%%%%%%%%%%%%%%%%%fig%%%%%%%%%%%%%%%%%%%%%%%%%%%%%%%
%
%
The T matrix may then be evaluated from
\be
\bl
&T(\om)
=
\\
&
\frac{1}{\De(\om)}
\left[
\begin{array}{cc}
V_+[1-V_-g_d(\om)]& V_+V_-g_c(\om)\\
V_+V_-g_c(\om)&V_-[1-V_+g_d(\om)]
\end{array}
\right]\\
\label{eq:Tmat2}
\el
\ee
with the determinant in the denominator given by
\bea
\bl
&\De(\om)=
\\
&
[1-V_+g_d(\om)][1-V_-g_d(\om)]-V_+V_-g_c(\om)^2
.\;
\label{eq:respos1}
\el
\eea
{\BLU We note that because the impurity scattering is momentum independent the T matrix depends only on the frequency. Therefore
the momentum structure for the T matrix expression in Eq.~(\ref{eq:Greencorr3}) is the same as in Born approximation (Eq.(\ref{eq:Greencorr1})) but the amplitudes of the corresponding QPI spectra may differ.}
Inserting the T matrix into Eq.~(\ref{eq:Greencorr3}) and taking the imaginary part 
$\de\rho(\omega)=-(1/\pi)ImTr^{(2)}\de G(\omega+i\eta)$ (where the double trace $Tr^{(2)}$ runs
over all matrix elements)  leads to the frequency dependent
spectral function correction at the impurity site that contains the effect of bound state  formation and concomitantly a modification of the continuum spectrum. The former leads to a pronounced peak below the conduction band edge which takes its weight from a reduced continuum spectral intensity. The explicit
expression for numerical evaluation of $\de\rho(\omega)$ obtained from Eqs.~(\ref{eq:Greencorr1}),
may be given in compact form in the case of scalar scattering $V_\pm=V_s$, we finally obtain for the spectral correction at the impurity site:
\bea
\delta\rho(\omega)=-\frac{2}{\pi}
{\rm Im}
\frac{V_sg(\om)^2}{1-V_sg(\om)}=\delta\rho_s(\omega)+\delta\rho_b(\omega),
\eea
where $\om=\omega +i\eta$ the definition $g(\om)=g_d(\om)+g_c(\om)$ was used. We have to consider two contributions to this spectrum: 
For frequencies lying inside the continuum of reconstructed bands (Figs.~\ref{fig:dispplot},\ref{fig:DOSplot})  
$g"(\omega)=-\pi(\fs\rho_0(\omega)+\rho_h(\omega))$ will be nonzero and we obtain explicitly
\be
\delta\rho_s(\omega)=-\frac{2V_s}{\pi}\Bigl(\frac{V_s (g^{'2}-g^{"2})+2g'g"(1-V_sg')}{(1-V_sg')^2+(V_sg")^2}\Bigr)_\omega.
\ee
This correction increases continuously with scattering strength $V_s$.  A more dramatic possibility is the appearance of a bound state below or anti-bound state above the band continuum for larger absolute values of $|V_s|$. Outside the continuum $g"(\om)\equiv 0$ and therefore the imaginary part determining $\delta\rho_b(\omega)$ originates only from the bound state pole of the T matrix which leads to a delta-function peak below the band bottom for {\it attractive} impurity interaction $V_s<0$:
\bea
\bl
\delta\rho_b(\omega)&=-\frac{2}{\pi}g'(\omega)^2
{\rm Im}
\frac{V_s}{1-V_sg(\omega)-i\eta}
\\
&
=A(\omega_b)\delta(\omega-\omega_b),
%&=&2g'(\omega_b)^2\Bigl |\frac{dg'}{d\omega}\Bigr |_{\omega_b}\delta(\omega-\omega_b)
\el
\eea
where $\omega_b$ is the bound state pole which appears below the band bottom and $A(\omega_b)$ is its weight given by
\bea
A(\omega_b)
&=&2g'(\omega_b)^2\Bigl (\int\frac{\fs\rho_t(\omega')}{(\omega_b-\omega')^2}\Bigr )^{-1},
\label{eq:bweight}
\eea
where $\rho_t(\omega)=\sum_\bq\rho_t(\bq,\omega)=\rho_0(\omega')+2\rho_h(\omega')$ (Eq.~(\ref{eq:totDOS1})) is the total local reconstructed DOS. The analogous possible appearance of the anti-bound state above the
top of the band for repulsive $V_s>0$ will not be considered here. The position of the bound state is given by the vanishing denominator in the above equation, explicitly by solving
%
% %%%%%%%%%%%%%%%%%%%%% figure %%%%%%%%%%%%%%%%%%%%%%%%%%%%
\begin{figure}
\includegraphics[width=0.99\columnwidth]{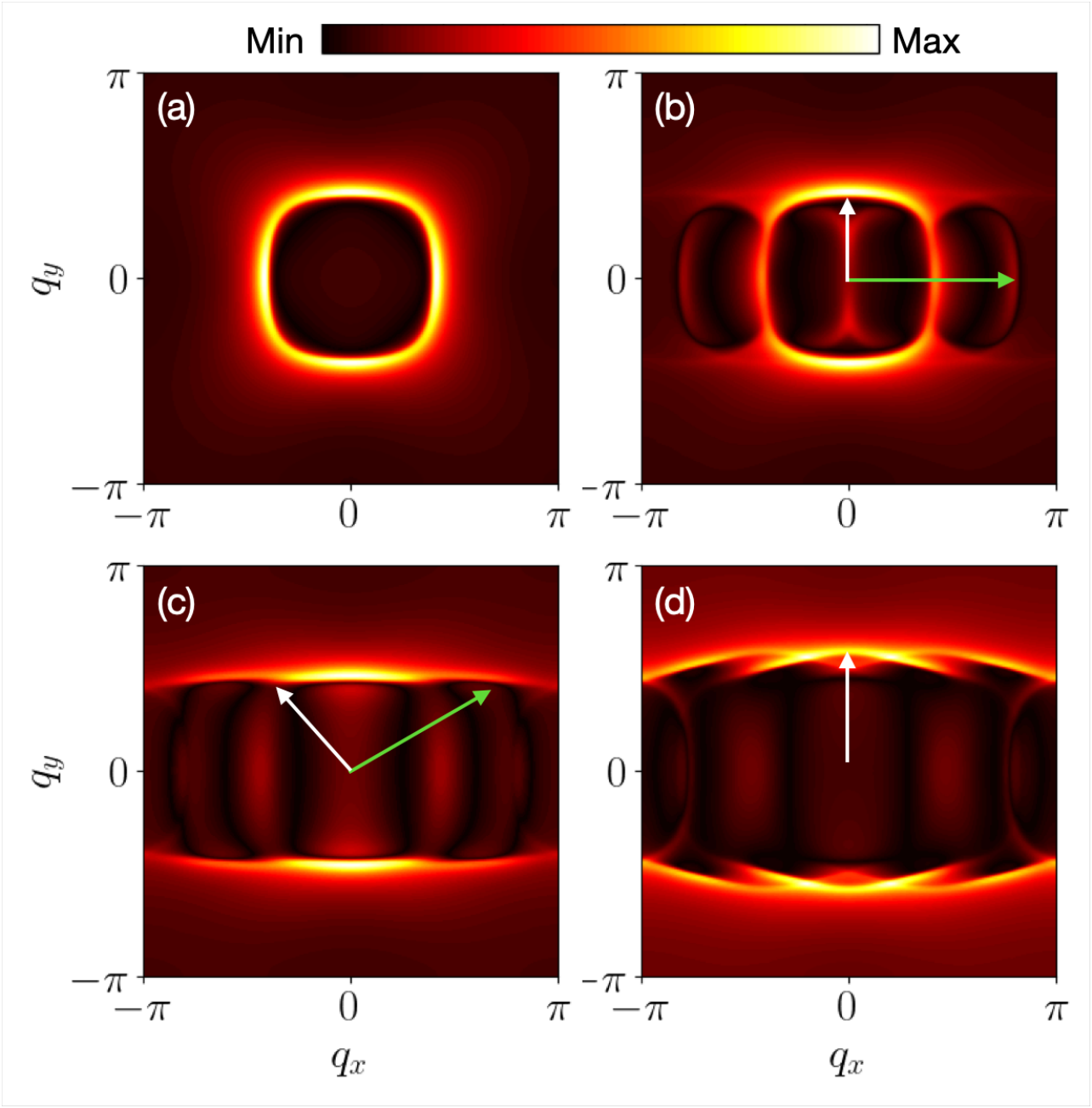}
\caption{ Total QPI spectra $\delta\rho_t(\hat{\omega}=0,\bq)=\delta_0+\delta_h$ (modulus) according to Eq.~(\ref{eq:delrho}) for various exchange coupling strengths $\gamma =0., 0.05, 0.25 ,0.5$ from (a)-(d). The Fermi surface model of Fig.~\ref{fig:FSpara}(a) with $\bQ=(0.4\pi,0)$  is used.  The increasing prominence of satellite images shifted by $\pm\bQ$ due to anomalous part $\rho_h$  with inreasing $\gamma$ and the reduction  of QPI  features from the gapped BZ regions are visible.
\label{fig:QPI_gavar}}
\end{figure}
%%%%%%%%%%%%%%%%%%%%%%fig%%%%%%%%%%%%%%%%%%%%%%%%%%%%%%%
%
 %
\bea
\frac{1}{V_s}=g'(\omega_b) =\int\frac{\fs\rho_t(\omega')}{\omega_b-\omega'}.
\label{eq:bpos}
\eea
Via the anomalous helical spectral density $\rho_h(\omega)$ contained in $\rho_t(\omega)$  both position and weight of the impurity bound state will depend on the coupling strength  $\gamma$ to the helical
ordered moments. The above equations for $\omega_b(V_s,\gamma)$ and $A(V_s,\ga)$ then
have to be solved numerically with the DOS functions of ,e.g., Fig.~\ref{fig:DOSplot}. This is illustrated in Fig.~\ref{fig:ispec}. It is instructive
to consider first the case of bare DOS with $\gamma=0$ i.e. $\rho_t(\omega)=\rho_0(\omega)=\rho_c(\omega)$. 
When we further move the zero energy position to the band center and simply approximate $\rho_c(\omega)$ by
a constant $(\bar{\rho}_0=1/2D)$ with $\omega\in[-D,D]$ lying between band bottom and top we can obtain analytical results for bound state position and weight $\omega_b^0$ and $A_0(\omega_b^0)$ as given by
\bea
\bl
\omega_b^0&=
-(D+\eps_b) = D\coth\frac{\alpha}{2},
\\
 A(\omega_b^0)
 &=\frac{1}{2D^2}(\omega_b^{02}-D^2)
\ln^2\bigl(\frac{\omega_b^0-D}{\omega_b^0+D}\bigr).
\label{eq:purebound}
\el
\eea
Here $\alpha=1/(\rho_0V_s)<0$ is a dimensionless parameter for the impurity potential. A bound state with $\omega_b^0<-D$ (or $\eps_b > 0$) exists
in the whole interval $-\infty <\alpha<0$ equivalent to $0<\rho_0|V_s|<\infty$.\\
The asymptotic behavior of these  approximate solutions for $\gamma=0$ is discussed in the Appendix \ref{sec:app-asymp}. For the general helical case with finite exchange coupling $\gamma$ the bound state pole has to be obtained from the numerical solution of Eqs.~(\ref{eq:bpos},\ref{eq:bweight}) which contains the true $\gamma$-dependent spectral function $\rho_t(\omega')$. The result is shown
 in Figs.~\ref{fig:ispec},\ref{fig:bound-avs} and discussed further below.

\section{Discussion of impurity QPI and bound state spectrum}
\label{sec:discussion}

In a metal without a symmetry breaking order parameter such as superconductivity, charge or 
spin density wave the images of QPI spectra at various bias voltages or frequencies $\hat{\omega}=eV=\omega-\eps_F$  contain
information on the equal-energy surfaces $\epsilon_\bk=\omega$ of the underlying conduction electron band structure. At low frequency
it may give an image of the Fermi surface in simple 2D cases  For example a 2D Fermi surface circle with radius
$k_F$ is mapped to a circle with radius $2k_F$ in the QPI image resulting from the impurity induced scattering wave vector \bq~ across the Fermi surface with a modulus $|\bq|=2k_F$. 

More complicated Fermi surfaces are usually not imaged as a whole by QPI but only certain characteristic
wave vectors $\bq_i$ that connect locally parallel sheets or cusps of the Fermi surface that have an enhanced
Fourier amplitude $\delta\rho(\omega,\bq_i)$ of the QPI density oscillations are seen. This can be used to map the Fermi surface, e.g. of the 2D electronic structure in doped cuprates \cite{zeng:24} or the 2D topological surface states in topological insulators \cite{lee:09,thalmeier:20,ruessmann:21}
%
% %%%%%%%%%%%%%%%%%%%%% figure %%%%%%%%%%%%%%%%%%%%%%%%%%%%
\begin{figure}
\includegraphics[width=0.8\columnwidth]{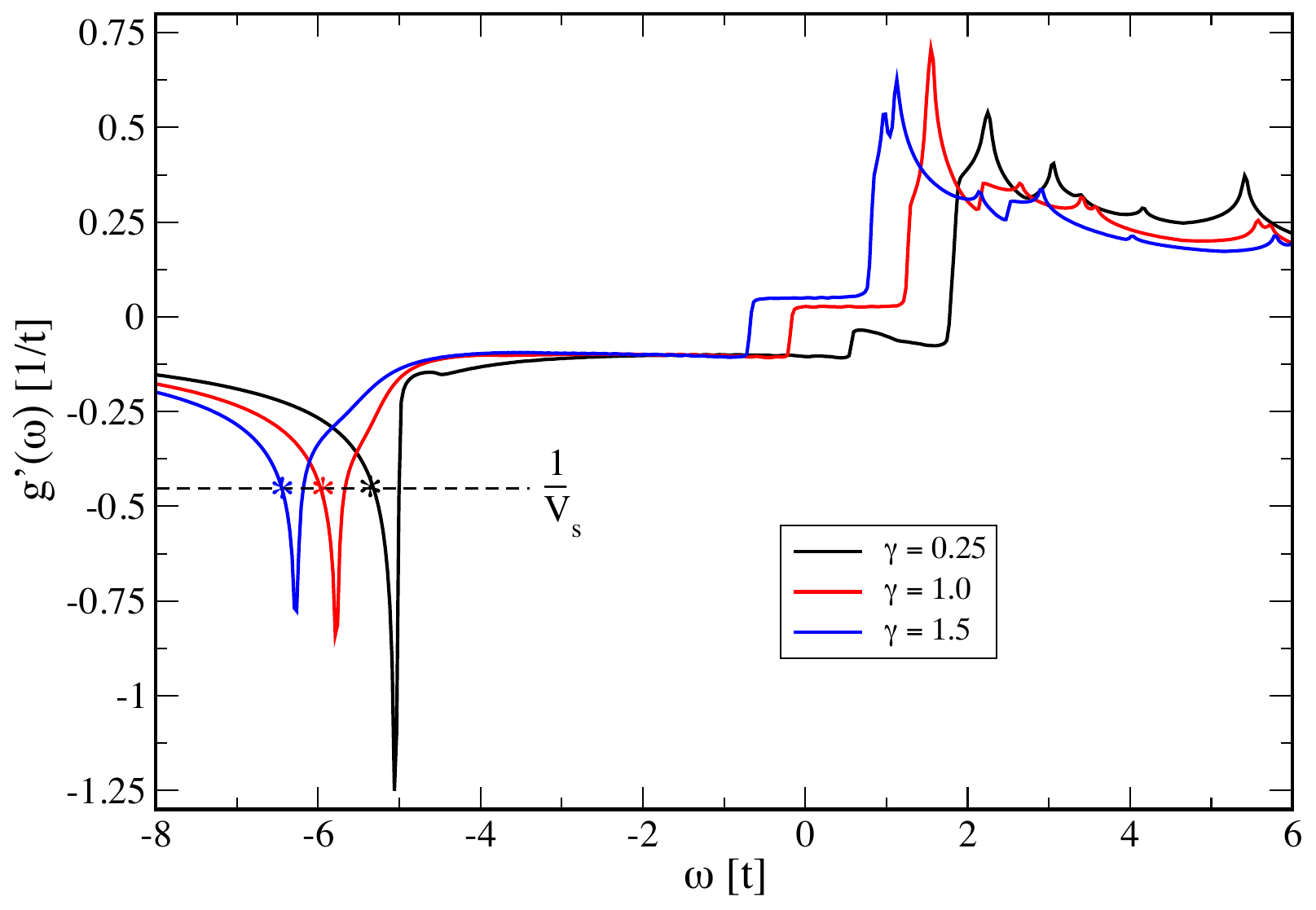}
\caption{Illustration of graphical solution for Eq.(\ref{eq:bpos}). The stars indicate bound state positions $\omega_b$ for various exchange couplings $\gamma$. They shift to lower energies with the lower band edges (the downward peaks) but the distance of $\omega_b$ to the edges changes little, mostly determined by the fixed $t/V_s=0.45$. }
\label{fig:boundplot}
\end{figure}
%%%%%%%%%%%%%%%%%%%%%%fig%%%%%%%%%%%%%%%%%%%%%%%%%%%%%%%
% %%%%%%%%%%%%%%%%%%%%% figure %%%%%%%%%%%%%%%%%%%%%%%%%%%%
\begin{figure}
\includegraphics[width=0.8\columnwidth]{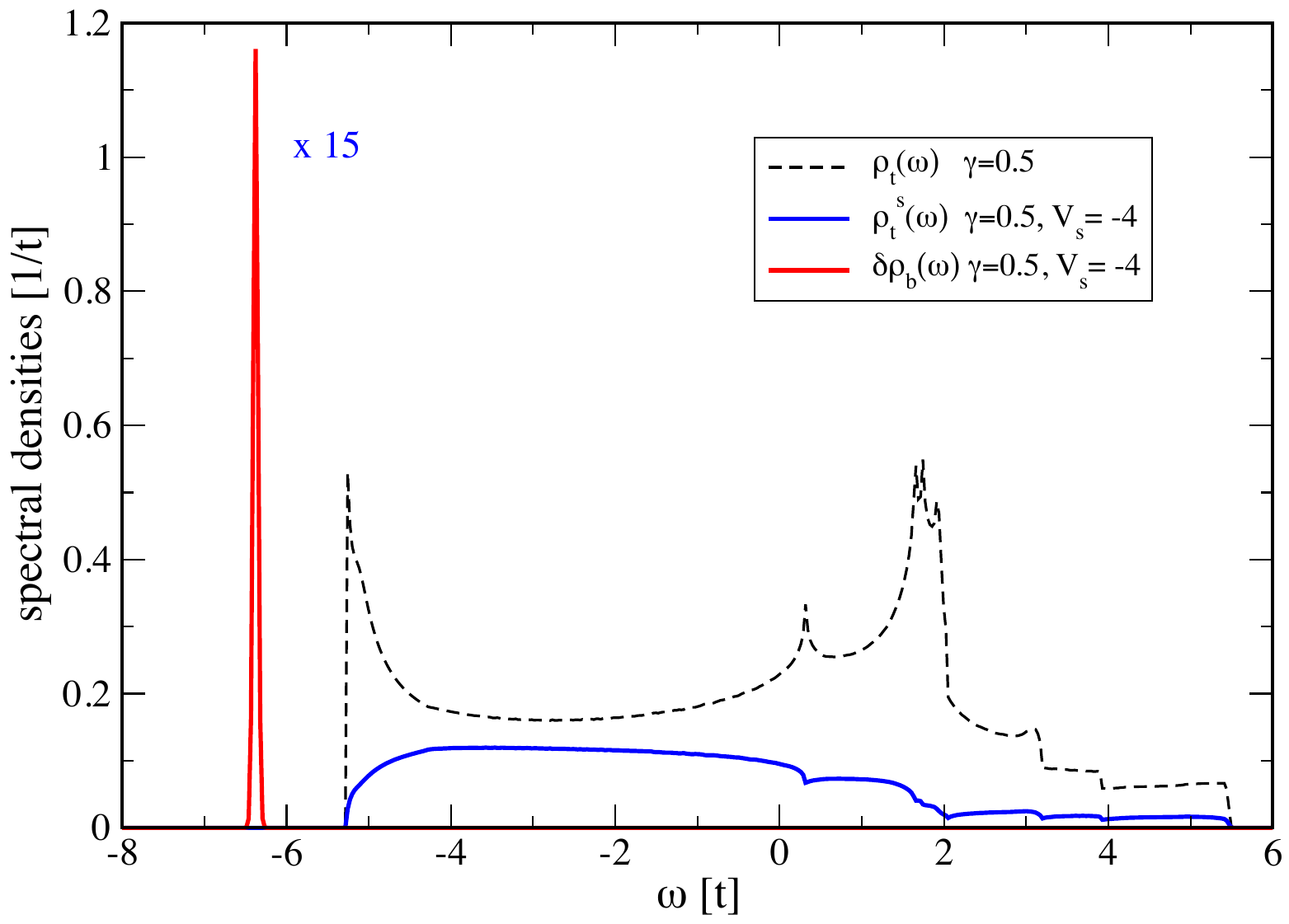}
\caption{Spectral densities for helical pure case ($\rho_t$), for the continuous spectrum $\rho_t^s$ at impurity site (the sum of $\rho_t$ and $\delta\rho$) and the impurity bound state delta-peak $\delta\rho_b$. The latter has been broadened and scaled down (by a factor 15) for presentation. The integrated intensity of the pure and impurity spectrum $\rho_s+\delta\rho_b$ is always equal to 2, independent of scattering strength $V_s$.  }
\label{fig:ispec}
\end{figure}
%%%%%%%%%%%%%%%%%%%%%%fig%%%%%%%%%%%%%%%%%%%%%%%%%%%%%%%
%
In this section we first discuss some of the aspects that appear in the QPI images of Figs.~\ref{fig:QPI_om0},\ref{fig:QPI_omvar},\ref{fig:QPI_gavar} that are due to the conduction bands reconstructed by localised helical magnetic background order. Naturally the interpretation is less straightforward in this case as may be seen from the  total QPI spectrum $\delta\rho(\omega,\bq)$ given by Eq.~(\ref{eq:delrho}). It contains {\it two} parts: The regular part $\delta\rho_0(\omega,\bq)$ where the wave vector \bq~of the QPI image connects reconstructed conduction states with  momenta \bk~and $\bk+\bq$, equivalent to conventional QPI  and a second anomalous part which is due to scattering connecting states with momenta \bk~and $\bk+\bq\pm\bQ$, therefore each wave vector \bq~in its Fourier transform has contributions from {\it three} scattering possibilities between states with \bk~and $\bk+\bq$ or $\bk+\bq\pm\bQ$ whereas in the normal part of QPI only the first of these exists.\\

The structure and appearance of QPI images is rather insensitive to the values of scalar and exchange scattering parameters. Therefore in all QPI figures we choose $V_+=V_-$, i.e. only nonzero scalar scattering $V_s$ which is then just an amplitude scale
for the images.
A comparative selection is shown in Fig.~\ref{fig:QPI_om0} for the three FS models of Fig.\ref{fig:FSpara} corresponding to the three rows. The first column again presents the spectral functions of  Fig.~\ref{fig:specpara} for easy  cross reference of characteristic scattering vectors that appear in the QPI images. The white and green arrows correspond  to scattering where the momentum transfer $\bq$ is determined by impurity scattering alone or where the SDW background order 
leads to a total transfer $\bq\pm \bQ$, respectively. According to  Eq.~(\ref{eq:delrho}) the former appear in normal as well as anomalous part of the total spectrum $\delta\rho_0$ and  $\delta\rho_h$ while the latter contribute only to the anomalous part $\delta\rho_h$ and this is the way how they are identified.
Therefore in  Fig.~\ref{fig:QPI_om0} we first present them separately. The second row for comparison shows
the QPI image of $\delta\rho\equiv\delta\rho_0$ without exchange coupling $(\gamma =0)$ which basically images the bare conduction band Fermi surfaces. There is no anomalous QPI contribution $\delta\rho_h$, therefore the scattering vectors that contribute are only from the first type (presented in white) and defined in the spectral function plots in the first column.

The third and last column in Fig.~\ref{fig:QPI_om0} represent $\delta\rho_0$ and $\delta\rho_h$ (their absolute values) for finite $\gamma/t=0.25$, respectively. Here we can directly see two effects of the coupling to helical magnetic order: Firstly the reconstruction of conduction bands leads to the vanishing or reduction of intensity of parts of the Fermi surface due to the gapping of states connected by the nesting vectors. This is particularly visible in the FS model of the first row. Secondly in the anomalous QPI spectrum (last column) additional image structures appear which are obviously shifted by $\pm\bQ$ from those in the regular QPI spectrum contribution 
of the third column. This is in accordance with the expressions for the two contributions in Eq.~(\ref{eq:delrho}). This effect is most prominent for
the first two FS examples with incommensurate $\bQ$. However we notice there is not much qualitative difference between regular and anomalous contribution for the almost perfect nesting case (bottom row). This is because adding or subtracting the commensurate $\bQ=(\pi,\pi)$ in $\rho_h$ leads to equivalent scattering vectors $\bq$ as in $\rho_0$.

The total experimentally observable QPI spectrum, however, consists of the sum of the two terms. Its absolute value
is shown in Fig.~\ref{fig:QPI_omvar}, now for fixed $\gamma/t=0.25$ and varying frequency (bias voltage). For zero or moderate
frequency the features and prominent scattering wave vectors  of both regular and anomalous contributions shown separately in Fig.~\ref{fig:QPI_om0} before are now both visible in the total spectrum. For larger frequencies the QPI image changes strongly
because the spectral functions are now different from those at $\hat{\omega}=0$  (first column in Fig.~\ref{fig:QPI_om0}).
{\BLU We note that in particular the QPI images of the upper row corresponding to FS model of Fig.\ref{fig:FSpara}(a) show strong
$C_4$ symmetry breaking due to the choice of a domain with $\bQ=(0.4\pi,0)$ ordering vector. For the orthogonal domain the image would simply be rotated by $\pi/2$.}

Finally in Fig.~\ref{fig:QPI_gavar} we show the evolution of the total QPI image at $\hat{\omega} =0$ (as in Fig.~\ref{fig:QPI_om0}) with increasing exchange coupling $\gamma$ for the FS model in Fig.\ref{fig:FSpara}(a).  It demonstrates that the features resulting from the anomalous contribution $\rho_h$ become more prominent when $\gamma$ increases as expected. And it shows that
the QPI features  connected by the nesting vector $\bQ=(0.4\pi,0)$ are strongly reduced in intensity due to the gapping and associated band reconstruction.\\
%
% %%%%%%%%%%%%%%%%%%%%% figure %%%%%%%%%%%%%%%%%%%%%%%%%%%%
\begin{figure}
\includegraphics[width=0.8\columnwidth]{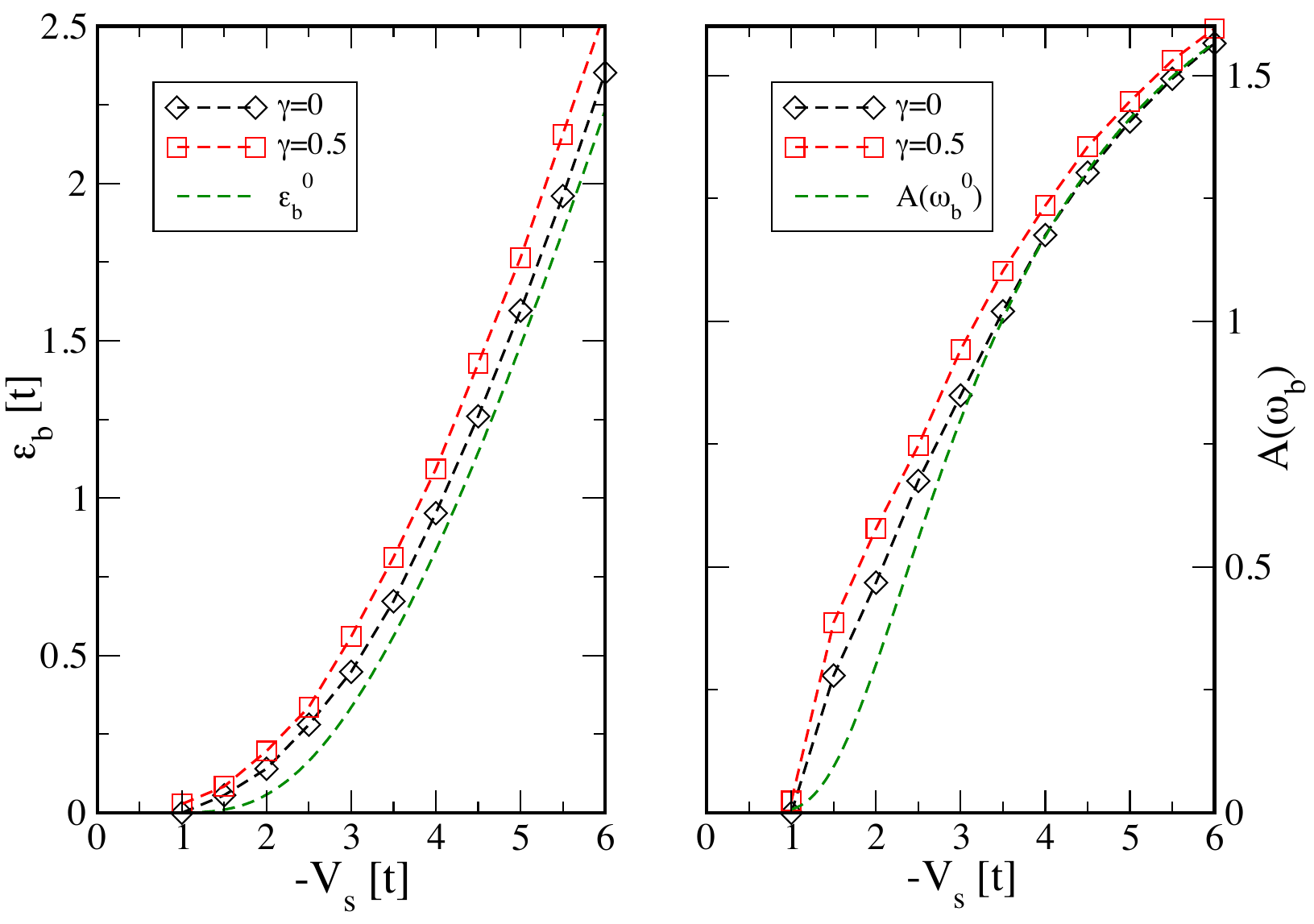}
\caption{Left panel: Dependence of boundstate position $\omega_b=-(D+\eps_b)$ (Eq.~(\ref{eq:bpos})) below the band  edge $-D$, with binding energy $\eps_b>0$, on the impurity scattering strength, corresponding to the FS model (a) in Fig.\ref{fig:FSpara}. It is shown for zero and finite exchange coupling $\gamma$  to helical  moment background. Right: similar plot for bound state weight $A(\omega_b)$ (Eq.~(\ref{eq:bweight})). The dashed green line corresponds to analytical solution for $\gamma=0$ with assumed constant $\bar{\rho}_0=1/2D$ (Eq.~\ref{eq:purebound})). }
\label{fig:bound-avs}
\end{figure}
%%%%%%%%%%%%%%%%%%%%%%fig%%%%%%%%%%%%%%%%%%%%%%%%%%%%%%%
%
In summary we can say that the modification of the impurity QPI spectrum due to presence of helical background magnetic order in a way mimics the behavior of the pure spectrum (Fig.\ref{fig:satellites}) as already mentioned below Eq.~(\ref{eq:rhofunc}).
QPI features visible at certain characteristic scattering vectors $\bq$ obtain satellites at positions $\bq\pm\bQ$. In the pure surface case, however, we have only a delta-peak at  $\bq=0$ which then obtains two easily identified satellites in Fig.\ref{fig:satellites}. In the QPI spectrum on the other hand the regular part $\rho_0$ in principle has already intensity at {\it every} wave vector $\bq$. Then adding two anomalous $\pm\bQ$ satellites from $\rho_h$ for every $\bq$ leads to a superposition of the BZ images in the total spectrum that may not easily or uniquely be separated.\\

Now we discuss the results for impurity  bound state formation within T matrix theory, including the effect of background helical magnetic order. The position of the bound state is given by the integral equation Eq.(\ref{eq:bpos})) and its weight by Eq.~(\ref{eq:bweight}). A graphical illustration for the solution of the {\BLU \it bound state position} corresponding to the first FS model of Fig.~\ref{fig:FSpara}(a) is shown in Fig.~\ref{fig:boundplot}. The bound state energies indicated by the stars are located below the band edge which corresponds to the sharp cusps above the stars. It shows that with increasing exchange coupling $\gamma$ both band edge and bound state energy shift to lower values. In concordance
with the bound state appearance the impurity-site continuum spectrum above the band edge is strongly modified. This can be seen in Fig.~\ref{fig:ispec} in a comparison of the case for $\gamma/t=0.5$ and strong scattering $V_s/t=4$ (red and blue curves) with that for absent exchange coupling ($\gamma=0$, dashed black) in the pure case without scattering. The continuum spectrum of the latter which is equal to $\rho_c(\omega)$ will be strongly reduced (blue curve) when $V_s$  is turned on and the missing  intensity will be shifted to the bound sate pole (red curve in {\BLU Fig.~\ref{fig:ispec}}, artificially broadened and reduced in peak height for presentation). {\BLU Its intensity is obtained by numerical integration in Eq.~(\ref{eq:bweight}).} The total intensity will always integrate up to two, the number of states per site. 
%
% %%%%%%%%%%%%%%%%%%%%% figure %%%%%%%%%%%%%%%%%%%%%%%%%%%%
\begin{figure}
\includegraphics[width=0.9\columnwidth]{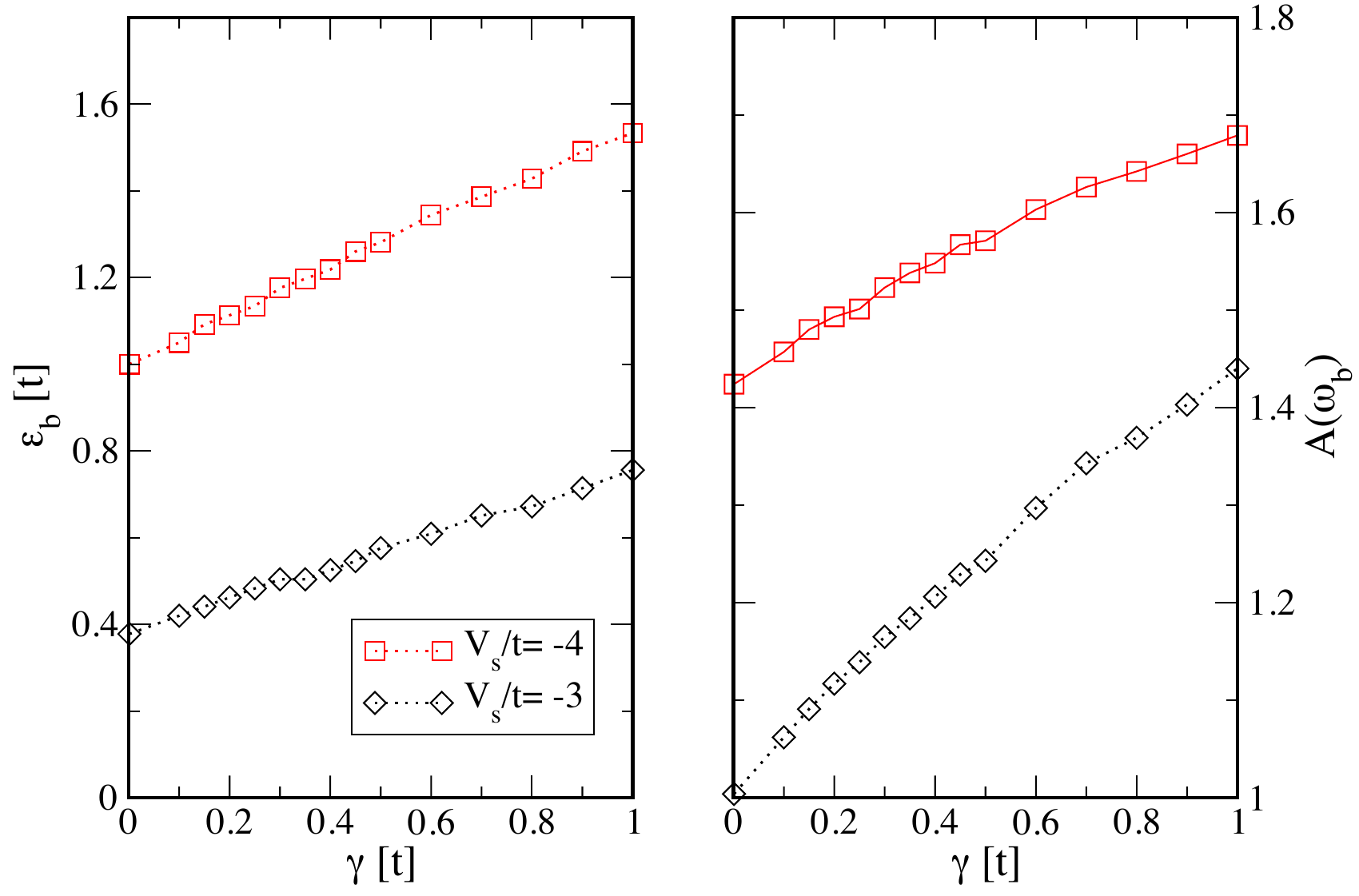}
\caption{Left panel: Dependence of binding energy $\epsilon_b$ on the exchange coupling strength
$\gamma=\fs SI_{ex}$ of conduction states to helical local moments for the FS model (e) in Fig.\ref{fig:FSpara},
close to perfect nesting case. In this FS model the (almost linear) increase with $\gamma$ is considerable due to the
complete FS gapping caused by helical order. Right panel: Corresponding changes in the bound state weight. }
\label{fig:bound-ga}
\end{figure}
%%%%%%%%%%%%%%%%%%%%%%fig%%%%%%%%%%%%%%%%%%%%%%%%%%%%%%%
%

The position of the bound state peak depends on the strength of impurity scattering and  the exchange coupling $\gamma$,
to the magnetic background. The evolution of its binding energy $\epsilon_b=-D-\omega_b>0$, i.e. the distance to the lower band edge $(-D)$ with $V_s$ and $\gamma$  is shown in Fig.~\ref{fig:bound-avs} (left panel). While the dependence on $V_s$ is strong $\epsilon_b$ is only moderately sensitive to $\gamma$ for the FS model of Fig.~\ref{fig:FSpara}(a).
This can be understood from Eq.~(\ref{eq:bpos}). The main effect of $\gamma$ seen in Fig.~\ref{fig:DOSplot} is a moderate local redistribution of weights in $\rho_t(\omega)$ in the continuum range. However, in Eq.~(\ref{eq:bpos}) the integration over $\omega'$ averages over these redistributions so that in the end the integral depends only weakly on $\gamma$. This is naturally also observed for the bound state weight given by Eq.~(\ref{eq:bweight}) which is presented in  Fig.~\ref{fig:bound-avs} (right panel) for this FS model.

For comparison we also show the dependence of binding energy and bound state weight for the almost perfect nesting FS model in Fig.~\ref{fig:FSpara}(e). The singular vanHove type DOS $\rho_c(\omega)$ (around $\omega\simeq 0$) will be fully gapped at this energy already by a small exchange coupling $\gamma$. This will lead to a major redistribution of spectral density and hence to a stronger modification of  the bound state integral in Eq.~(\ref{eq:bpos}). Therefore the dependence of binding energy $\eps_b(\gamma)$ and bound state weight $A(\gamma)$ on the exchange coupling is more pronounced as in the previous case. This is shown in Fig~\ref{fig:bound-ga} for two impurity scattering strengths $V_s$.

\section{Summary and Conclusion}
\label{sec:summary}

In this work we  developed a theory for the STM spectrum of a two-component system, consisting of conduction electrons
and localised electrons whose magnetic moments form a helical (generally incommensurate) structure. The two subsystems
are coupled by a contact exchange interaction. The conduction electrons are described by a tight binding model and for consistency the local moment helix propagation vector \bQ~is taken as the main nesting vector of the resulting conduction band Fermi surface as  witnessed by the maximum in the Lindhard function.

The motivation and starting point of the investigation is the observation that local moment helix order is seen in the compound \GR~with stable $S=7/2$ Gd moments. Since the 4f electrons in a stable 4f shell cannot significantly contribute to the tunneling current the visibility of helix order in the STM spectrum of this compound \cite{spethmann:24,yasui:20} must be due to an indirect effect caused by the exchange coupling of the two subsystems.\\

Our theoretical description is based on the real space Green's function approach, for pure surfaces as well as under the presence of surface impurities. For the pure surface the background helical order induces satellite peaks at positions $\pm\bQ$  in Fourier transform of the conduction electron density in addition to the main central peak. For inhomogeneous periodic conduction electron density each reciprocal lattice vector $\bK_s$ acquires satellites at $\bK_s\pm\bQ$. In this way the STM pure surface topographic mapping reveals
the magnetic ordering vector of the localised electrons which do not directly contribute to the tunneling current.\\

A related effect is observed in the quasiparticle interference spectrum caused by impurity scattering. In Born approximation the resulting nonperiodic conduction electron charge modulations are characterized by the sum of a regular and an anomalous helical term . The former
contains contributions from impurity scattering with momentum transfer $\bq$ whereas the latter has combined contributions with momentum transfer $\bq$ as well as $\bq\pm\bQ$. This leads to multiple appearing QPI features for each of the characteristic wave vectors $\bq_i$ that connect critical points or parallel sheets of the underlying Fermi surface. Furthermore certain parts of the QPI image related to the nesting or ordering vector may be gapped out or 
reduced in intensity for moderate frequencies.
The QPI spectrum under the presence of helical order is therefore more complicated to interpret than in the nonmagnetic case but on the other hand through its additional structures gives information on the reconstruction of the conduction bands by the helical magnetism.

Finally the strong impurity scattering case was treated with full T matrix theory and shown to lead to impurity bound state formation. The position
of bound states below the continuum depends strongly on the scattering potential. The influence of the exchange coupling to helical order depends considerably on the Fermi surface model, it is strongest for the nearly perfect nesting case. Provided the amplitude of the helical moments can be varied, e.g. by applied field or temperature variation its influence on the bound state formation may be observable in STM experiments with high energy resolution.

{\BLU The principal effects of helical order investigated here have been motivated by their observation in \GR. They may
well be present in other magnetic $4f$-122 compounds if suitable surfaces can be prepared. Another possibility
to study the influence of magnetic order on STM experiments is provided by the single component Fe-pnictides \cite{kamble:16,goyal:25,Akbari:10a} where
the magnetism is directly appearing as a SDW in the conduction electron system. }

\appendix

\section{Fourier transform of pure surface densities for cell periodic Bloch states}
\label{sec:app-cellper}

In the main text we have derived the spectral properties using plane wave conduction states.
For the background spectral densities of the pure surface this leads to the regular central peak at $\bk=0$ and 
to satellite peaks at $\pm\bQ$. Here we show that under the assumption of general periodic
Bloch wave functions for conduction states the satellites may  appear around any reciprocal
lattice vector $\bK_s$, i.e. at positions $\bK_s\pm\bQ$. The spinors in Eq.~(\ref{eq:Greendef})
$\Psi^\dag(\br')=(\psi^*_{\bk\ua}(\br'),\psi^*_{\bk+\bQ\da}(\br'))$ then contain the  Bloch states
$\psi_{\bk\si}(\br)=(1/\sqrt{N})exp(i\bk\cdot\br)u_\bk(\br)\chi_\si$ where $u_\bk(\br)=u_\bk(\br+\bR_i)$
are the cell-periodic functions with $\bR_i$ denoting a lattice vector. For a single conduction band they
may be taken as real and can be expanded in a Fourier series according to
\bea
u_\bk(\br)=\sum_s b_\bk(\bK_s)e^{i\bK_s\br}.
\label{eq:cellfunc}
\eea
Then, using the Green's function for helical reconstructed states in Eq.~(\ref{eq:Greendef})
we obtain
\be
\bl
&&G_0(\br,\br,\omega)=\frac{1}{N}\sum_\bk\{ [u^2_\bk(\br)g_a(\bk,\omega)+u^2_{\bk+\bQ}(\br)g_b(\bk,\omega)]
\\
&&+g_c(\bk,\omega)u_\bk(\br)u_{\bk+\bQ}(\br)(e^{i\bQ\cdot\br}+e^{-i\bQ\cdot\br})\},
\label{eq:purelocalin}
\el
\ee
which is the generalisation of the plane wave result Eq.~(\ref{eq:purelocal}) to cell periodic Bloch states. The products
of the cell periodic functions $u_\bk(\br)$ entering the above equation have the same periodicity and may also be
expanded in a Fourier series. Denoting the Fourier coefficients by $B_\bk(\bK_t)$ $B_{\bk+\bQ}(\bK_t)$ and $A_{\bk,\bQ}(\bK_t)$ for the 
three terms, respectively, taking the Fourier transform and subsequently the imaginary part of the Green's function we
finally arrive at the total spectral density
\be
\bl
&\rho_t(\omega,\bq)=\rho_0(\omega,\bq)+\rho_h(\omega,\bq)=
\\[0.3cm]
&\sum_s\{\frac{1}{N}\sum_\bk[B_\bk(\bK_s)R_\ua(\bk,\omega)+B_{\bk+\bQ}(\bK_s)R_\da(\bk,\omega)]\delta_{\bq,\bK_s}
\\
&+\frac{1}{N}\sum_\bk A_{\bk\bQ}(\bK_s)R_h(\bk,\omega)(\delta_{\bq,\bK_s+\bQ}+\delta_{\bq,\bK_s-\bQ})\},
\label{eq:totDOSin}
\el
\ee
where regular and anomalous spectral functions  $R_{\ua\da}(\bk,\omega)$ and  $R_{h}(\bk,\omega)$ are defined in Eqs.~(\ref{eq:regDOS},\ref{eq:helDOS}). This constitutes the generalisation of Eq.~(\ref{eq:totDOS1}) for the plane wave case to the cell-periodic Bloch functions. The first part is the regular spectral density with peaks at  reciprocal lattice vectors $\bK_s$. The second part of this equation demonstrates that the satellite peaks of the total helical  reconstructed  spectral density of the pure surface now appear around every reciprocal lattice vector, i.e., at $\bq=\bK_s\pm\bQ$ and not only at $\bq=\pm\bQ$ as in the plane wave case. Finally we give the coefficients in the above equation in terms of the fundamental Fourier coefficients of the cell periodic functions in Eq.~(\ref{eq:cellfunc}) as convolutions
\be
\bl
B_\bk(\bK_s)&=\sum_{s'}b_\bk(\bK_{s'}-\bK_s)b_\bk(\bK_{s'}),
\\
A_{\bk\bQ}(\bK_s)&=\sum_{s'}b_{\bk+\bQ}(\bK_{s'}-\bK_s)b_\bk(\bK_{s'}).
\el
\ee
In the plane wave case this reduces to $B_\bk(\bK_s)=A_{\bk\bQ}(\bK_s)=\delta_{\bK_s,0}$  and accordingly Eq.~(\ref{eq:totDOSin}) leads
to the previous result of Eq.~(\ref{eq:totDOS1}).

\section{Asymptotic behavior of bound state solutions}
\label{sec:app-asymp}

To understand qualitatively the impurity bound state properties it is instructive to consider asymptotic limits of the simplified model  of Eq.~(\ref{eq:purebound}) for the unreconstructed bands using the rough approximation
of a constant $\bar{\rho}_0=1/2D$.\\

i) {\it weak impurity scattering}: $\omega_b^0=-D-\epsilon_b$ close to lower band edge with $\epsilon_b\ll D$. This corresponds to large $|\alpha|$ or small $|V_s|$. 
Then we have for bound state position and weight
\be
\epsilon_b\simeq 2D\exp(-|\alpha|);\;\;\;  A(\epsilon_b)=\frac{\epsilon_b}{D}\ln^2\frac{2D}{\epsilon_b}.
\ee
As the bound state approaches the lower band edge $(\epsilon_b\rightarrow 0$) its weight vanishes.\\

ii) {\it strong impurity scattering}: $\omega_b^0$ far removed from lower band edge $\omega_b^0\ll -D$ for $\alpha\rightarrow 0^-$ or very large $|V_s|$.
In this somewhat unphysical limit we get
\be
\omega_b^0\simeq -\frac{2D}{|\alpha|}=-|V_s|;\;\;\; A(\omega_b^0)\simeq 2.
\ee
Here the bound state has absorbed all the weight from the continuum states.
In other words at the impurity site the continuum spectral density vanishes whereas all the intensity goes into
the bound state pole.

%\begin{acknowledgments}
%\end{acknowledgments}

%\clearpage

%%%%%%%%%%%%%%%%%%%%%%%%      References        %%%%%%%%%%%%%%%%%%%%
%\newpage
%\bibliographystyle{prsty}
\bibliography{References}
\end{document}